\documentclass[aps,prl,reprint]{revtex4-2}

\usepackage{graphicx}
\usepackage{amsmath,amssymb,amsfonts}
\usepackage{bm} 
\usepackage{hyperref} 
\usepackage{float} 
\usepackage{color}
\usepackage{booktabs}
\usepackage{subcaption}
\usepackage[utf8]{inputenc}

\usepackage{multirow}
\usepackage{algorithm}
\usepackage{algorithmicx}
\usepackage{algpseudocode}
\usepackage{listings}
\usepackage{appendix}
\usepackage{caption}
\begin{document}

\title[Article Title]{Cyclic Quantum Annealing: Searching for Deep Low-Energy States in 5000-Qubit Spin Glass}

\author{Hao Zhang*$^{1}$, Kelly Boothby$^{2}$ and Alex Kamenev$^{1,3}$}

\affiliation{$^1$School of Physics and Astronomy, University of Minnesota, Minneapolis, MN 55455, USA}
\affiliation{$^2$D-Wave Systems Inc., Burnaby, British Columbia, Canada}
\affiliation{$^3$William I. Fine Theoretical Physics Institute, University of Minnesota, Minneapolis, MN 55455, USA}

\begin{abstract}
Quantum computers promise a qualitative speedup in solving a broad spectrum of practical optimization problems. The latter can be mapped onto the task of finding low-energy states of spin glasses, which is known to be exceedingly difficult. Using D-Wave's 5000-qubit quantum processor, we demonstrate that a recently proposed iterative cyclic quantum annealing algorithm can find deep low-energy states in record time. We also find intricate structures in a low-energy landscape of spin glasses, such as a power-law distribution of connected clusters with a small surface energy.  These observations offer guidance for further improvement of the optimization algorithms.     
\end{abstract}

\maketitle

\section{Introduction}\label{sec1}

The concept of a quantum computer, as originally envisioned by Feynman\cite{feynmanSimulatingPhysicsComputers1982a}, is a system consisting of a large number of spins. With the advent of quantum devices encompassing more than 5000 qubits in spin-like system\cite{brookeQuantumAnnealingDisordered1999b,johnsonQuantumAnnealingManufactured2011,dicksonThermallyAssistedQuantum2013a,kingObservationTopologicalPhenomena2018,kingCoherentQuantumAnnealing2022,mohseniIsingMachinesHardware2022,kingQuantumCriticalDynamics2023}, quantum computing has gained even more substantial momentum, showing promise across various fields\cite{perdomo-ortizFindingLowenergyConformations2012, harrisPhaseTransitionsProgrammable2018, mottSolvingHiggsOptimization2017b,kingQuantumAnnealingSimulation2021, abelQuantumFieldTheoreticSimulationPlatform2021}, including the exploration of dynamics of quantum phase transition\cite{kingQuantumCriticalDynamics2023}.

One of the most promising applications is the approximate solving of a broad spectrum of discrete optimization problems. Many such problems can be formulated in terms of finding low-energy  states of spin systems\cite{barahonaComputationalComplexityIsing1982,farhiQuantumAdiabaticEvolution2001,battagliaOptimizationQuantumAnnealing2005,lucasIsingFormulationsMany2014}. Quantum annealing uses quantum fluctuations to navigate the system between local minima towards its  low-energy states. 
It has been designed, developed, and validated as an effective method for an approximate solution of these optimization tasks\cite{kirkpatrickOptimizationSimulatedAnnealing1983,finnilaQuantumAnnealingNew1994,kadowakiQuantumAnnealingTransverse1998a,brookeQuantumAnnealingDisordered1999b,farhiQuantumAdiabaticEvolution2001,santoroTheoryQuantumAnnealing2002a,dasColloquiumQuantumAnnealing2008a,moritaMathematicalFoundationQuantum2008,youngFirstOrderPhaseTransition2010,albashAdiabaticQuantumComputation2018b,ohkuwaReverseAnnealingFully2018a,yamashiroDynamicsReverseAnnealing2019,chancellor_experimental_2021,haukePerspectivesQuantumAnnealing2020a,passarelliReverseQuantumAnnealing2020,rajakQuantumAnnealingOverview2023}. Ideally, the ground state is attainable if the adiabatic condition is strictly met\cite{ santoroTOPICALREVIEWOptimization2006, moritaMathematicalFoundationQuantum2008,dasColloquiumQuantumAnnealing2008a,albashAdiabaticQuantumComputation2018b}. However, the spin-glass phase\cite{edwardsTheorySpinGlasses1975,binderSpinGlassesExperimental1986a,santoroTheoryQuantumAnnealing2002a,crisantiComplexitySherringtonKirkpatrickModel2003,cavagnaNumericalStudyMetastable2004,mukherjeePossibleErgodicnonergodicRegions2018,harrisPhaseTransitionsProgrammable2018} presents significant challenges, characterized by a vast number of local minima and exponentially small energy gaps, which require exponential time to satisfy adiabaticity. Any deviations from  adiabaticity lead to a cascade of Landau-Zener transitions\cite{zenerNonadiabaticCrossingEnergy1932,sinitsynMultiparticleLandauZenerProblem2002, santoroTheoryQuantumAnnealing2002a, volkovExactResultsSurvival2004, damskiSimplestQuantumModel2005, sinitsynSolvableMultistateModel2016,wangManybodyLocalizationEnables2022} to higher energy states, making deep low-energy states improbable.

This raises a question: how to minimize Landau-Zener transitions and go deeper into the energy landscape of the spin glass. Recently, an iterative cyclic quantum annealing algorithm\cite{wangManybodyLocalizationEnables2022} has been proposed. In this paper we demonstrate that the key elements of this algorithm — the reference Hamiltonian\cite{perdomoStudyHeuristicGuesses2010, chancellorModernizingQuantumAnnealing2017, caoSpeedupQuantumAdiabatic2021, wangManybodyLocalizationEnables2022} and cycling through the many-body localization transition\cite{palManybodyLocalizationPhase2010a, gornyiManybodyDelocalizationTransition2016,mukherjeeManybodyLocalizationdelocalizationTransition2018} — are indeed useful steps towards improved quantum optimization routines. These elements, used together, allow for the manipulation of the spin-glass energy landscape, effectively reducing the number of Landau-Zener transitions and facilitating a deeper exploration of low-energy states.

\begin{figure}[H]
  \centering
  \includegraphics[width=\linewidth]{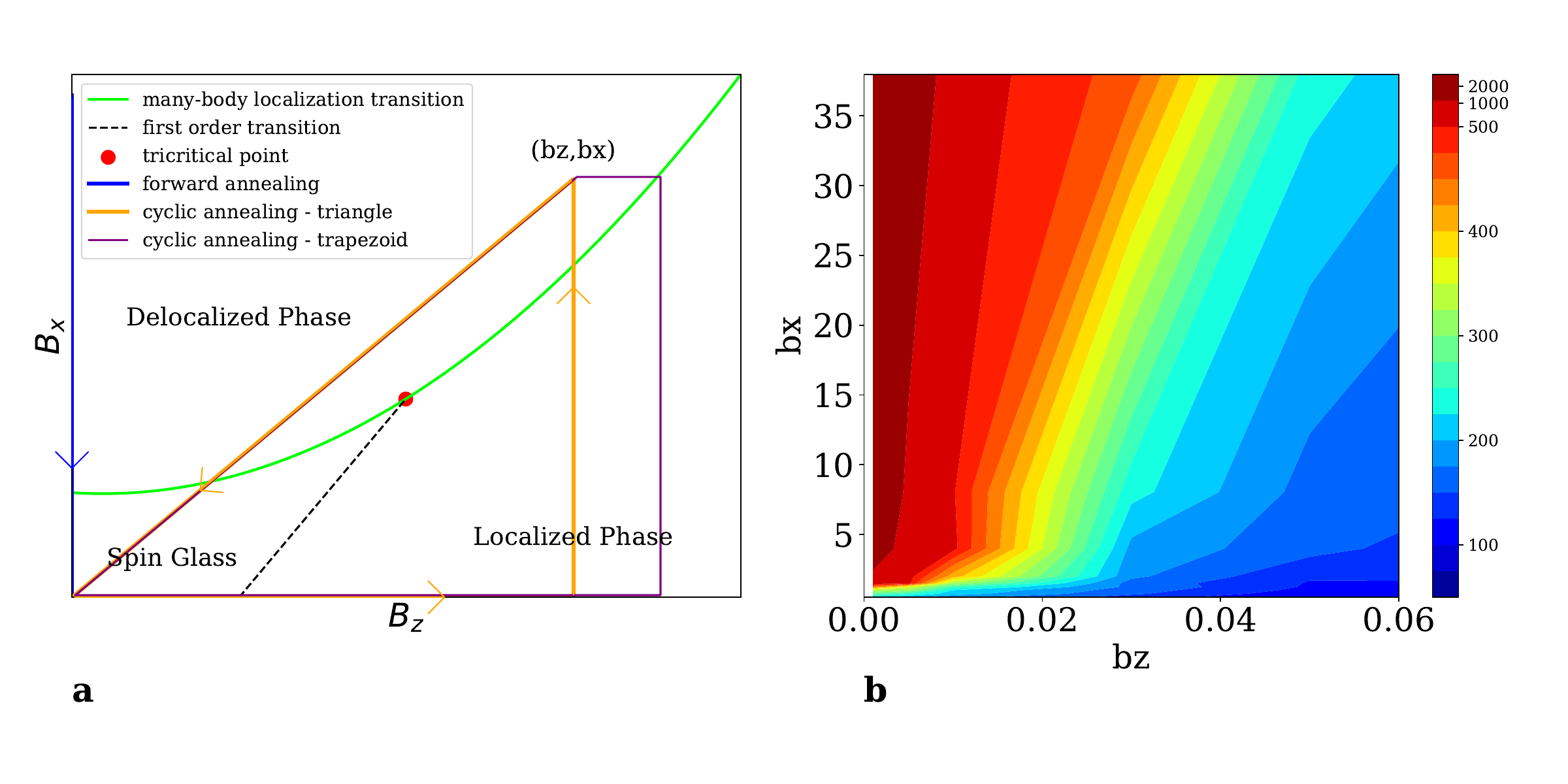}
  \caption{\textbf{a) Schematic Phase Diagram.} Presented here is a schematic representation of the phase diagram for the system under study, as defined in Eq.~\ref{eq:main}. The green line denotes the many-body localization transition, while the black dashed line indicates the first-order transition. The path of the cyclic annealing algorithm is represented by the orange triangle contrasting with the blue path, which illustrates the path taken by standard forward annealing. A trapezoid cycle for cyclic annealing is also shown here by the purple line. \textbf{b) Contour of Hamming Distance.} Cycles with different $(b_z,b_x)$ are sampled to produce an approximate phase diagram. The average Hamming distance between the final and initial states of a single cycle is calculated and depicted as a contour plot. The contour lines in Hamming distance may indicate the boundary of the many-body localization transition. The disparate magnitudes of $b_x$ and $b_z$ arise from the normalization of the problem Hamiltonian.}
  \label{fig:phase_diagram}
\end{figure}

Using D-Wave's 5000-qubit quantum processor, the Advantage\textsuperscript{\texttrademark} system, to simulate large spin glasses\cite{kingQuantumCriticalDynamics2023}, we demonstrate that cyclic quantum annealing can indeed reach deep low-energy states in the spin-glass landscape and can surprisingly save 85 percent of the annealing time resources as compared to traditional forward annealing. Considering the limited availability of annealing time, cyclic quantum annealing demonstrates significant practical utility.

Analyzing the local energy minima found by the algorithm, we observed interesting structures in the low-energy landscape of the spin glasses. Spins flipped between relatively close local minima tend to cluster with a single large cluster contributing most to the energy difference. The distribution of cluster sizes follows a power-law distribution with an exponent close to -2.1, which shows that rare events (i.e., atypical clusters) predominantly contribute to energy balance. These findings 
provide useful clues into further improvements to optimization search. 

Cyclic quantum annealing can provide experimental data on the statistics of deep minima in large 3D spin glasses. This advancement has the potential to shed light on longstanding conjectures in the field, see, e.g.\cite{palassiniTrivialityGroundState1999,marinariEffectsChangingBoundary2000,marinariEffectsBulkPerturbation2001}.



\section{Cyclic Quantum Annealing}\label{subsec2}

The traditional forward  quantum annealing\cite{rajakQuantumAnnealingOverview2023} involves evolution along an open path that connects a Hamiltonian with ground state close to a product state with a Hamiltonian with unknown ground state.  The idea is illustrated by the blue path in Fig.~\ref{fig:phase_diagram}a.). Cyclic annealing\cite{wangManybodyLocalizationEnables2022} takes a closed path in the parameter space, passing through the Hamiltonian of the problem, depicted as an orange path in Fig.~\ref{fig:phase_diagram}a.). Such a cyclic process allows the energy to progressively decrease, cycle by cycle, until a satisfactorily deep low-energy state is achieved.

The algorithm operates based on the following Hamiltonian:
\begin{equation}
    H(s) = H_p+B_z(s)H_\text{ref}+B_x(s)H_q, \label{eq:main}
\end{equation}
where $s$ parameterizes the cycle. It comprises three components, each serving a distinct function and corresponding to different steps of the algorithm:

\textit{Step 1 - Prepare the problem Hamiltonian:}
\begin{equation}\label{eq: H_p}
    H_p=\sum_{ij}^N J_{ij}\, \sigma_i^z\sigma_j^z.
\end{equation}
The problem Hamiltonian $H_p$ is an Ising-like spin Hamiltonian with $\sigma_i^z$ representing the $z$-Pauli matrices, realized by qubits. Coupling parameters, $J_{ij}$,  encode information about the optimization problem at hand. All couplers $J_{ij}$ available in the quantum processor are used in the simulations described below. The topology of $J_{ij}$ is essentially a 3D lattice, more details can be found in \cite{boothby_next-generation_2020}. The initial qubit states for the first cycle can be chosen as an \textit{arbitrary} $z$ polarized product state.  We use the trivial all-up-spin state for this purpose. In the last step of each cycle, the initial qubit state for the next cycle is updated. This state is called the reference state: it is a $z$ polarized product state with components denoted as $s_i^r$.  

\textit{Step 2 - Turn on the reference Hamiltonian:}
\begin{equation}
  H_\text{ref}=-\sum_i^N s_i^r\sigma_i^z.   
\end{equation}
This term biases the problem's Hamiltonian towards the reference state $\{s_i^r\}$. 
This step is realized by increasing $B_z(s)$ from 0 to $b_z$, passing through the first-order transition. This transition marks the $B_z$ field where the reference state becomes a true ground state of  the Hamiltonian $H_p+B_z(s)H_\text{ref}$ . 


\textit{Step 3 - Turn on the quantum Hamiltonian:}
\begin{equation}
    H_q = -\sum_i^N\sigma_i^x.
    \label{eq:driving}
\end{equation}
This term, which is non-commuting with respect to $H_p$ and $H_\text{ref}$, 
makes the states to be superpositions of bit-string states by providing matrix elements between them. For sufficiently high $B_x$, it drives the system through the many-body localization transition.  This step is achieved by increasing $B_x(s)$ from 0 to $b_x$. Due to the bias induced in Step 2, the mixing is mainly with the states whose
$H_p$ energy is less than that of the reference state.

\textit{Step 4 - Complete the cycle and perform measurement:}
The final step is to turn off $H_\text{ref}$ and $H_q$, completing a full cycle by returning to the original problem Hamiltonian. Measurements taken in the end cause the state to collapse into a classical bit-string states. The turn-off process involves numerous Landau-Zener transitions, thus the final state is not the ground state. However, due to the bias $H_\text{ref}$, the final state has a considerable probability of being lower in energy than the initial reference state. Each cycle is repeatedly sampled multiple times in the quantum processor to obtain a distribution of states. The lowest-energy state of this distribution is then selected. If it is lower than the initial state of the current cycle, it becomes the new reference state for the next cycle; otherwise, the reference state does not change.

\section{Simulations}\label{section:performance}

\begin{figure*}[hbtp]
  \centering
  \includegraphics[width=\linewidth]{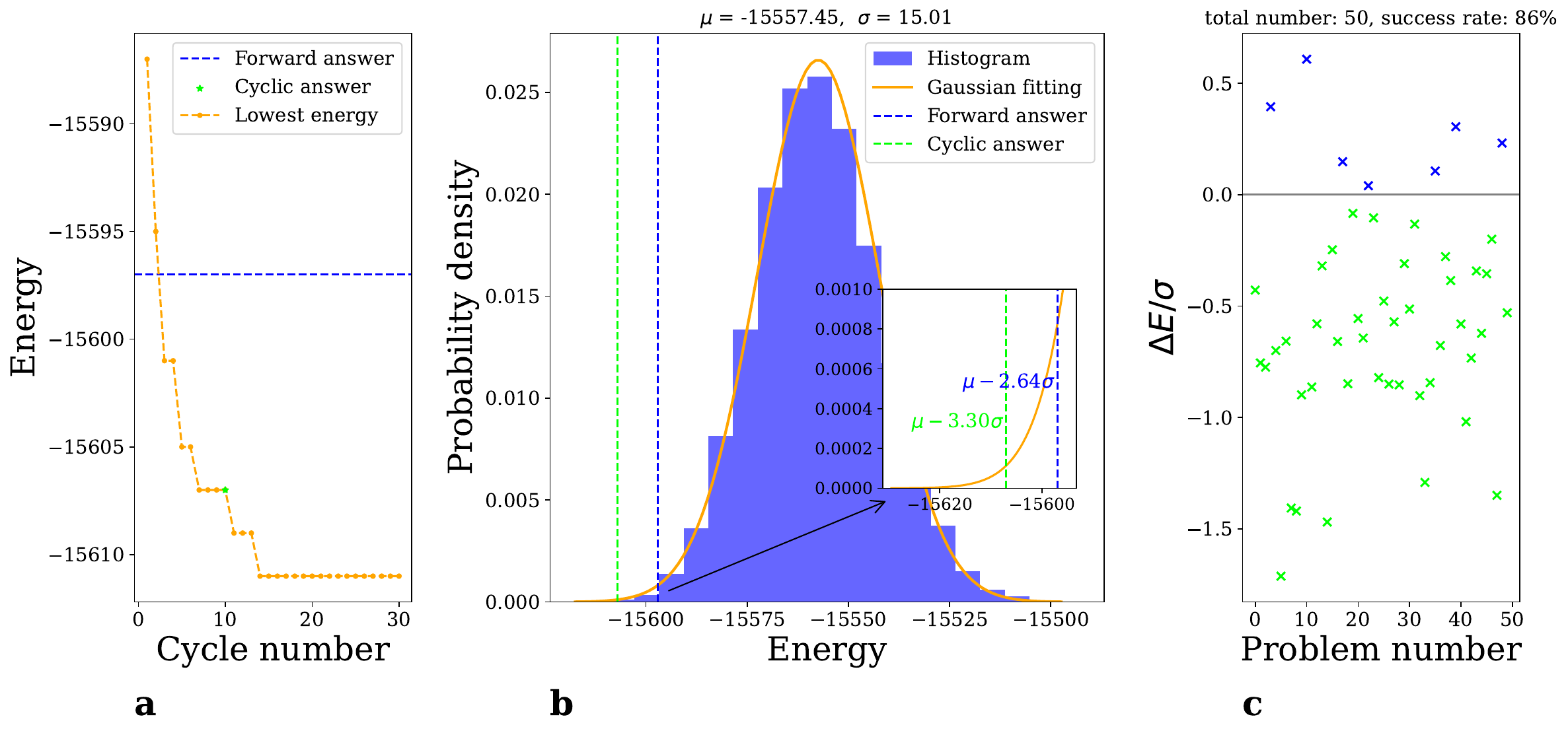}
  \caption{\textbf{Performance of Cyclic Annealing.} \textbf{a) Decreasing Energy Curve.} The orange dots represent the progressively lower energy state found by cyclic annealing for a selected problem from our comparison set.  Green stars and blue dashed lines mark the answers from cyclic annealing and forward annealing respectively. Notably, by the third cycle, cyclic annealing outperforms forward annealing. \textbf{b) Position in the Gaussian Distribution.} The histogram depicts the distribution of states obtained through forward annealing, fitting a Gaussian distribution with $\mu=-15685.47, \sigma=19.30$; see the orange line. The relative positions of cyclic (green dashed line) and forward (blue dashed line) annealing solutions in the distribution highlight the deeper reach of the former; see inset.  \textbf{c) Results of Comparison Experiment.} In an equal-time comparison of cyclic and forward annealing across 50 problems, the energy differences $\Delta E$ (normalized to the forward annealing’s standard deviation $\sigma$) are marked by crosses. Cyclic annealing's superiority is evident, with lower-energy solutions in $86\%$ of cases (green crosses).}
  \label{fig:Performance}
\end{figure*}

Multiple cycles of the cyclic quantum annealing algorithm are executed, to progressively find deeper low-energy states, as illustrated by the decreasing energy curve (orange line) in Fig.\ref{fig:Performance}a. More details can be found in the Methods section.

Evaluation of cyclic annealing performance involves direct comparison with forward annealing with the same amount of annealing time $T$, that is, cyclic annealing takes $T/N$ for each of the $N$ cycles. The latter typically produces results that fit a Gaussian distribution; see the histogram and orange line in Fig.\ref{fig:Performance}b. For a specific problem, we run it 1000 times and take the lowest energy as solution. This is marked by a blue dashed line in Fig.\ref{fig:Performance}a and b. In contrast, cyclic annealing is executed for 10 cycles and takes the answer of last cycle as solution (illustrated by the green star in Fig.\ref{fig:Performance}a and the green dashed line in Fig.\ref{fig:Performance}b). The cycle is the triangle with parameter $b_z=0.03, b_x=8.04$; see Fig.\ref{fig:phase_diagram}a.

The comparative study encompassed a set of problems chosen to represent typical spin-glass-type hard optimization problems in the form of Eq. \ref{eq: H_p}, ee the Methods section. The energy differences $\Delta E$ between the solutions obtained from cyclic annealing $E_\text{cyclic}^f$ and forward annealing $E_\text{forward}^f$, normalized by the standard deviation of the latter $\sigma$, are indicated by crosses. Notably, in $86\%$ of these comparisons, cyclic annealing achieved lower energy states (green crosses), affirming its enhanced performance, see Fig.\ref{fig:Performance}c. Regarding the instances where the cyclic approach did not succeed, this can be attributed to fluctuations inherent in forward annealing, which occasionally explores deeper low-energy states due to the tail behavior of its distribution. The average improvement in energy achieved by cyclic annealing is quantified as:
\begin{equation}
\overline{\left(\frac{\Delta E}{\sigma}\right)} = -0.56\pm0.49. 
\label{eq:sigma_difference}
\end{equation} Consequently,  solutions derived from cyclic annealing are positioned, averaging over the problem set, at a significantly lower energy level relative to the mean energy level of forward annealing (see green vertical line and blue vertical line in inset of Fig.\ref{fig:Performance}b for a typical problem), as given by:
\begin{equation}
\begin{split}
\overline{\left(\frac{E_\text{cyclic}^f - \mu}{\sigma}\right)} &= - 3.42 \pm 0.57 \\
\overline{\left(\frac{E_\text{forward}^f - \mu}{\sigma}\right)} &= - 2.86 \pm 0.40
\end{split}
\end{equation}
with $\mu$ representing the mean and $\sigma$ the standard deviation of the forward annealing distribution. From  Eq.~(\ref{eq:sigma_difference}) one can show that to achieve the same performance as cyclic annealing, forward annealing should be performed about 7000 times instead of 1000 times. In other words, cyclic annealing saves about 85 percent of the annealing time to achieve same performance as forward annealing. 

\begin{figure*}[hbtp]
  \centering
  \begin{subfigure}{0.85\textwidth}
    \includegraphics[width=\linewidth, height=0.5\textheight, keepaspectratio]{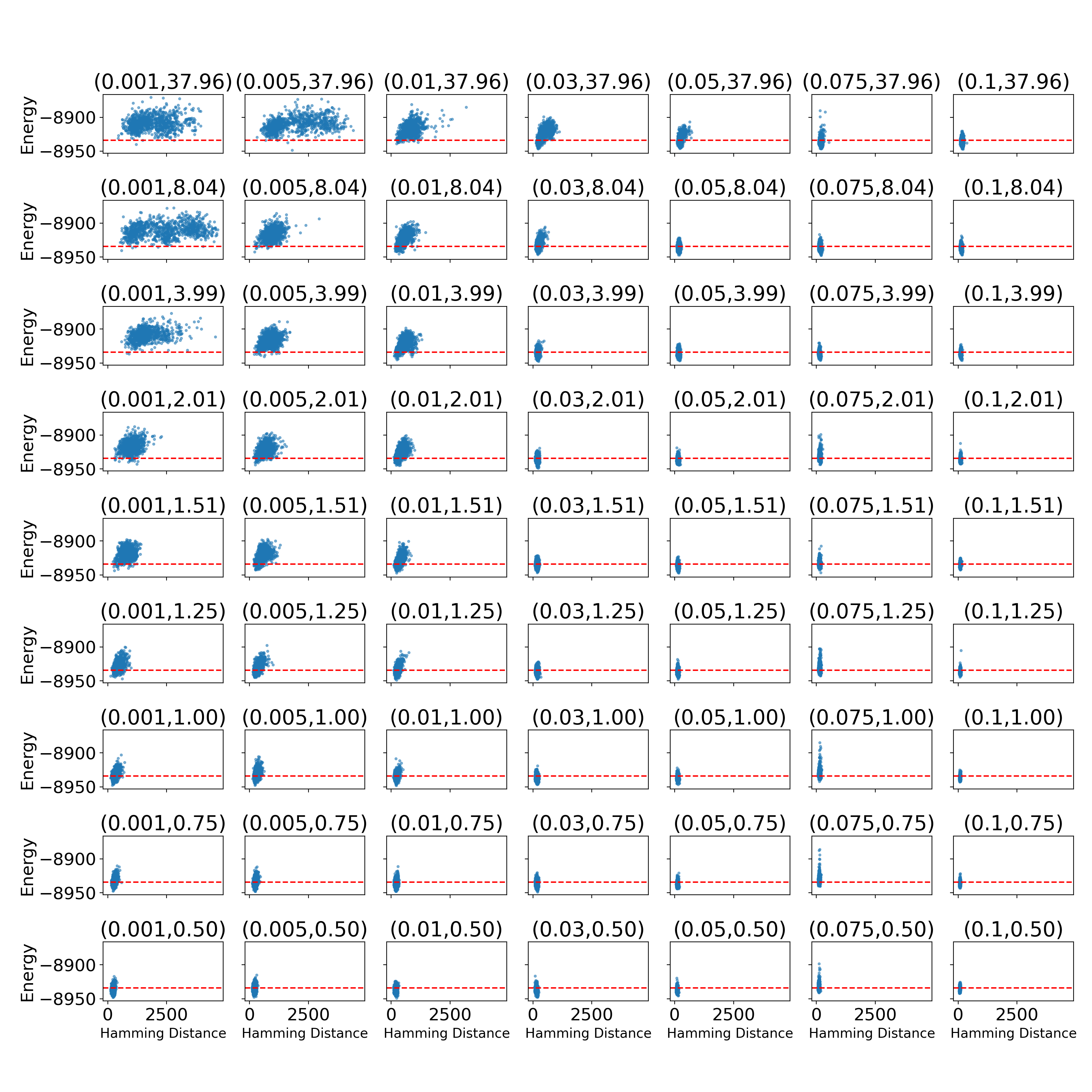}
    \caption*{\textbf{a}}
  \end{subfigure}
  \hfill
  \begin{subfigure}{0.85\textwidth}
    \includegraphics[width=\linewidth, height=0.3\textheight, keepaspectratio]{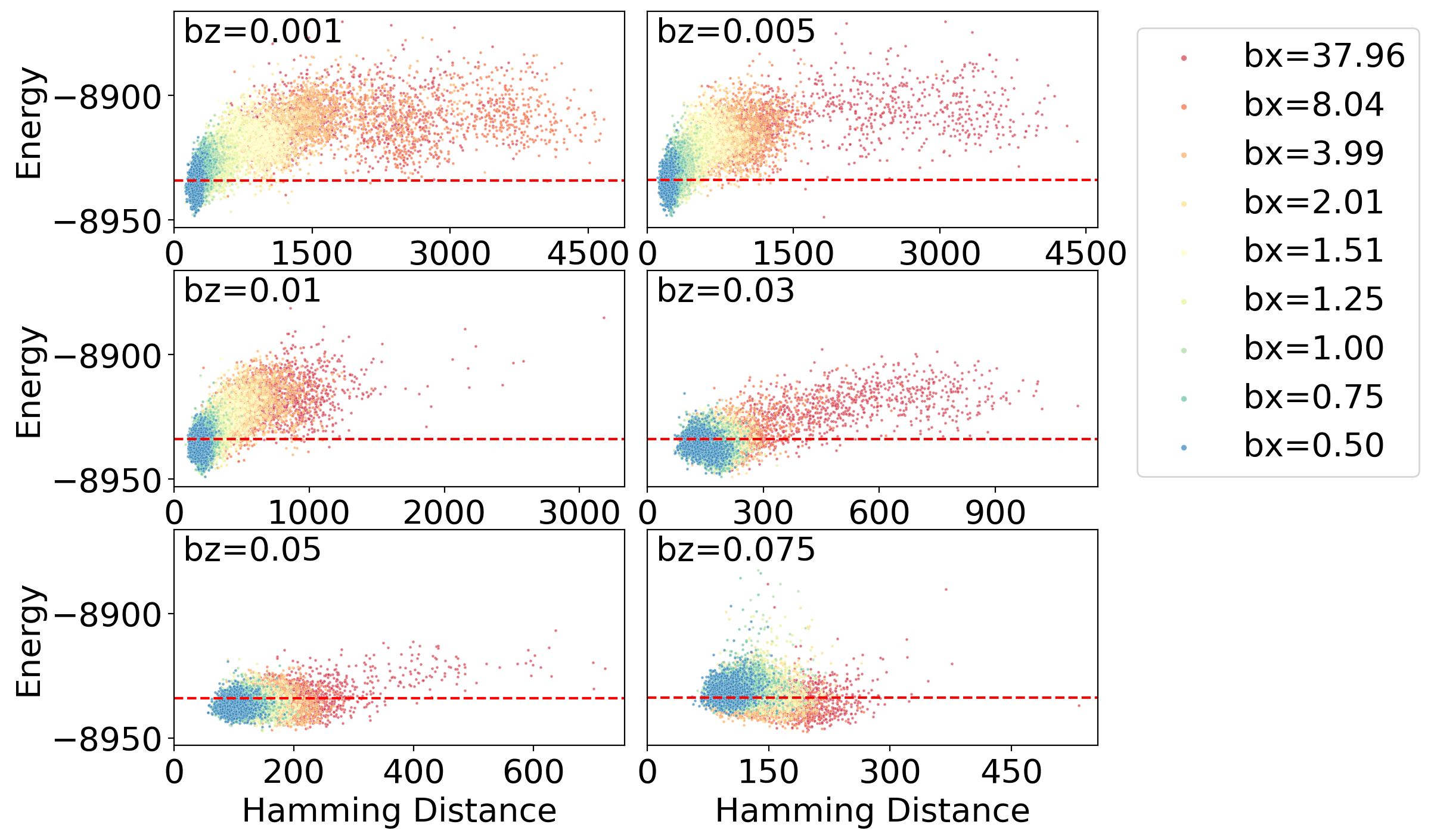}
    \caption*{\textbf{b}}
  \end{subfigure}
  \caption{\textbf{Distribution of Final States for a Single Cycle.} A wide range of the path parameters $(b_z,b_x)$ of trapezoidal cycle were sampled, see main text. The problem Hamiltonian, initial qubit states, and other parameters were kept fixed. For each data point, a single cycle is executed 1000 times to produce a distribution of final states. They are shown as dots in the plane of energy and Hamming distance, which is measured from the reference state.  Red dashed lines represent the energy of the reference state. \textbf{a) Overview.} Sub-figures are arranged according to $(b_z,b_x)$. \textbf{b) Enlarged view.} Distributions of varying $b_x$ but identical $b_z$ are shown in the same sub-figure. Six distinct $b_z$ values are selected.}
  \label{fig:distribution}
\end{figure*}

\section{Phase Diagram}\label{section:explore}

To better understand how cyclic annealing operates and to find the optimal parameters, we analyzed the distribution of final states reached by a single cycle. A trapezoidal cycle defined by the corners $(0,0)$, $(b_z',0)$, $(b_z',b_x)$, $(b_z, b_x)$ was used; see the purple cycle in Fig.~\ref{fig:phase_diagram}a.  To ensure the occurrence of the first-order transition in step 2,
$b_z'$ is set to a specific value. The choice of a trapezoidal path is intended to not miss the first-order transition. For larger values of $b_z$, this trapezoid can be simplified to a triangle. A broad range of path parameters $(b_z,b_x)$ were sampled with each fixed problem Hamiltonian.  A problem Hamiltonian is created by setting the $J_{ij}$ values for all available couplers to be uniformly distributed between $-1$ and $1$. For each data point, a single cycle is executed.  It is repeated 1000 times to generate a distribution of final states. The outcomes are shown as dots in the plane of energy and Hamming distance, which is measured from the reference state; see Fig.~\ref{fig:distribution}a and Fig.~\ref{fig:distribution}b.

These distributions show that cyclic annealing can explore a wide range of the Hamming distances for a 5000-qubit spin glass, from several hundreds to several thousands away from the reference state. For a fixed $b_x$, an increase in $b_z$ augments the magnitude of energy-spectrum distortion, thereby reducing the number of Landau-Zener transitions (as indicated by the shrinking distribution range).  This facilitates  a deeper exploration in the energy spectrum (evident from the lowering of the energy distribution). For a fixed $b_z$, a higher $b_x$ increases the magnitude of the many-body delocalization effects (visible from the increased Hamming distance distribution), albeit at the cost of elevating the energy of distribution. Taking these two competing effects into account, optimal parameter combinations can be chosen in practice; see enlarged view in Fig.\ref{fig:distribution}b.

Furthermore, we calculate the average Hamming distance and present it as a contour plot in Fig.\ref{fig:phase_diagram}b. The lines of equal hamming distance potentially hint at the boundary of the many-body localization transition. It is also noteworthy that an advanced variant of the cyclic annealing algorithm can be designed to, for example, dynamically adjust the $(b_z, b_x)$ parameters for each cycle.

\section{Discussion}

Cyclic quantum annealing can explore the structure of low-energy states in spin-glass systems. It follows the same approach as above for data acquisition. Cyclic annealing of a single cycle is performed with a variety of annealing parameters $b_x$, while maintaining $b_z$ at a constant value of 0.03. The reference state is chosen as a fixed low-energy state. In the analysis, each final state is compared to the reference state to identify flipped spins. These flipped spins are classified into clusters based on their connectivity via non-zero $J_{ij}$ couplings. A flipped spin $i$ is considered part of a cluster if it is connected to another flipped spin $j$ in the cluster by a nonzero coefficient $J_{ij}$.

\begin{figure}[hbtp]
  \centering
  \includegraphics[width=\linewidth]{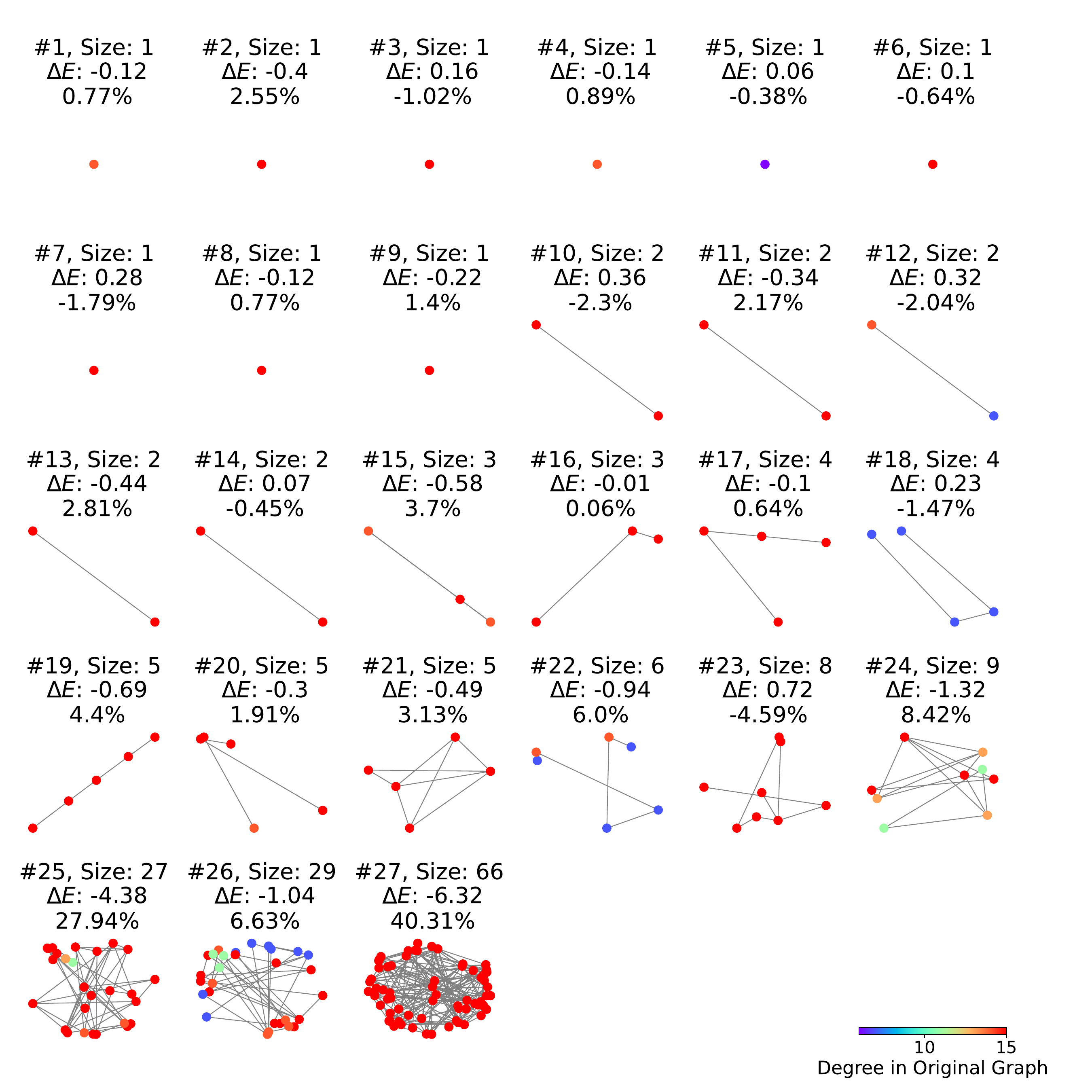}
  \caption{\textbf{Clusters of Flipped Spins in a Low-Energy State.} This figure illustrates the clusters formed by flipped spins when a low-energy state is compared to the reference state. Flipped spins are identified, and those connected by nonzero $J_{ij}$ values are grouped into the same cluster. The clusters thus identified are depicted here, with each cluster's size and contribution to the total energy difference (relative to the reference state) detailed alongside its percentage. Notably, a single, large cluster – Cluster 27 – emerges as the most significant contributor, accounting for 40.31\% of the total energy difference. The color coding represents the degree of each flipped spin in the original graph, which is constructed with nodes representing all spins and links denoting nonzero $J_{ij}$ connections.}
  \label{fig:cluster_visual}
\end{figure}

This analysis reveals a tendency for flipped spins to aggregate into clusters. Fig.~\ref{fig:cluster_visual}
illustrates the clusters for a typical low-energy state, showing the distribution and structure of all the flipped clusters. The size of a cluster is determined by the number of flipped spins it contains. The energy difference $\Delta E$ represents the contribution of a particular cluster $C$ of a state $\{s_i\}$ to the overall energy difference relative to the reference state $\{s_i^r\}$,
\begin{equation}
    \Delta E= \sum_{i\in\text{C}}\sum_{j\in N_i} J_{ij}(s_is_j-s_i^rs_j^r),
\end{equation}
where $N_i$ is the set of all nearest neighbor spins of $s_i$.

Analyzing multiple low-energy states, we consistently observe the presence of a dominant cluster that plays a substantial role in the overall energy difference. For instance, in the case illustrated in Fig.~\ref{fig:cluster_visual}, the largest identified cluster, Cluster 27, contributes a significant $40.31\%$ to the total energy difference. This highlights the critical impact of the largest clusters within the spin system, underscoring their importance in the overall energy landscape of the system.

\begin{figure}[hbtp]
  \centering
  \includegraphics[width=\linewidth]{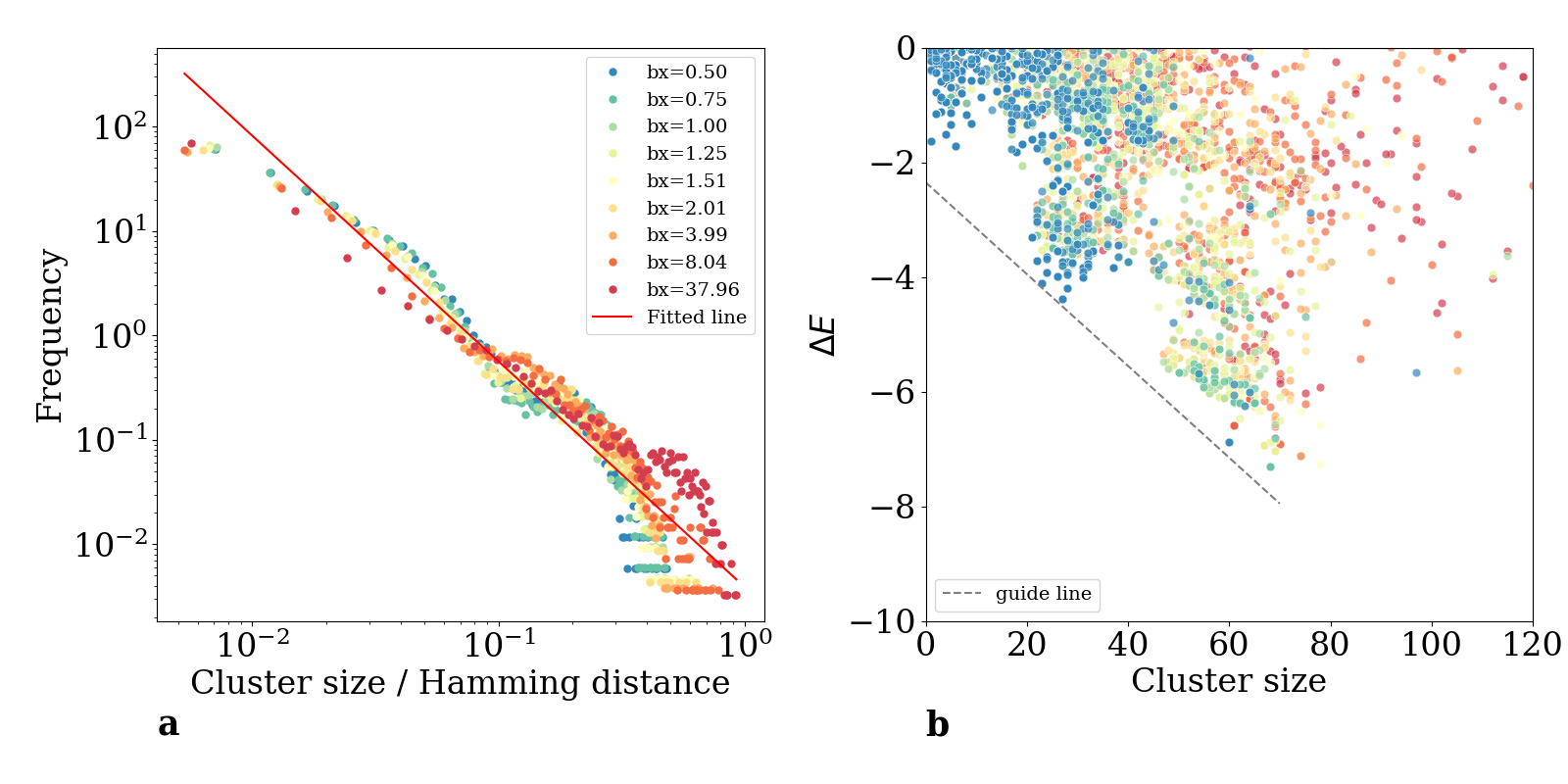}
  \caption{\textbf{Distribution of Clusters.} Clusters are identified for final states from cycles with different annealing parameters $b_x$. The parameter $b_z$ is fixed as $0.03$. \textbf{a) Power Law Distribution of Cluster Sizes.} Data collapse for clusters of final states from cycles with different $b_x$. The red line is fitted with power-law equation: $\log(y) = -2.12 \log(x) -0.03$ with r-value -0.95. The exponent $2.12$ shows the variance is not finite, implying the possibility of rare super clusters. \textbf{b) Energy Contribution $\Delta E$ vs. Cluster Sizes.} For each cluster, we calculated its energy contribution $\Delta E$ to the energy difference with respect to the reference state. Clusters from cycles with different annealing parameters $b_x$ are shown here in different color. The legend is the same as that in \textbf{a.} Data of negative $\Delta$E are shown here. A dashed line shows the relation between energy decrease and smallest possible cluster size to achieve that. To find a large energy decrease, one must find a large cluster. }
  \label{fig:cluster}
\end{figure}

Figure~\ref{fig:cluster}a
shows the frequency of finding a connected cluster of flipped spins of a given size for data from different $b_x$, which has same number of states and roughly same number of clusters.  It is plotted against the size of the cluster, normalized to the total Hamming distance, $d$, between the two low-energy states. For example, the normalized distance 1/2 means that the given cluster contains half of the spins flipped between the two states.  The probability of finding a cluster of size $s$  follows a power-law distribution $ P(s,d) \propto \left(\frac{s}{d}\right)^{-\alpha} $. Since all data with different $d$ collapse with same exponent, after summation over $d$, one get
\begin{equation}
    P(s) \propto s^{-\alpha},
\end{equation}
with the exponent $\alpha\approx 2.12$ from fitting.
This pattern emerges consistently across clusters identified in final states obtained from cycles with varying annealing parameters $b_x$. The data collapse observed for clusters from different $b_x$ values underscores this characteristic as a universal feature of the system. The fact the exponent $\alpha<3$ shows that the variance is not defined. This implies a prevalence of large clusters within the patterns of flipped spins, as mentioned above.

Figure~\ref{fig:cluster}b shows cluster's flip energy change, $\Delta E$, versus its size, $s$. It indicates a relationship between the magnitude of the energy reduction and the minimum size of the cluster required to achieve such a decrease. This shows that substantial energy decreases are possible upon the formation of larger clusters. 

In summary, transitions between local minima of the spin-glass landscape involve flipping a certain number of connected clusters. Cluster sizes follow an algebraic distribution with the exponent $\alpha\approx 2.12$. This makes large clusters rather prevalent. Moreover, there is a direct proportionality between the achieved energy gain and the cluster size. Targeting the larger clusters with negative energy may be the key to achieving significant energy reductions within the low energy part of the spin-glass landscape, as it may help to escape the local energy basin.  These observations have practical implications  for the development of optimization algorithms, which should be geared towards identifying and flipping large clusters. 

It is of considerable interest to investigate the performance of cyclic quantum annealing in comparison with the classical algorithms. Furthermore, it would be pertinent to explore the potential integration of the idea of cyclic quantum annealing into classical algorithms to improve their performance\cite{misra-spieldenner_mean-field_2023}.

\section{Methods}\label{sec4}

This section outlines the details of the cyclic quantum annealing algorithm and simulations. The algorithm is implemented on D-Wave's Advantage 4.1 quantum processor, which features 5,627 functional qubits and 40,279 couplers. We utilized all available qubits and couplers in our studies.

The problem set in the Simulations section consists of 50 random realizations of spin glass configurations: for Problem 0 to problem 9, $J_{ij}$ is randomly chosen as ±1; for problem 10 to problem 19,  $J_{ij}$ is randomly chosen as ±1, ±2; for problem 20 to problem 29, $J_{ij}$ is randomly chosen as ±1, ±2, ±3, ±4; for problem 30 to problem 39, $J_{ij}$ is randomly chosen as ±1, ±2, ±3, ±4 … ±8; for problem 40 to problem 49, $J_{ij}$ is randomly chosen as ±1, ±2, ±3, ±4 … ±16. Given that the underlying topology closely resembles a 3D lattice, these problems represent 3D spin glass systems, which are known to be NP-hard. We normalize the problem to have average $|J_{ij}|$=1 before the simulations.

For the cyclic annealing, the initial qubit state for the first cycle can be arbitrary $z$ polarized product state. Interestingly, starting with a high-energy state does not hurt the algorithm's efficiency, as it rapidly converges to a low-energy state after the first cycle.  An important aspect of the algorithm involves specifying the duration of each segment of the path. Table~\ref{table:path-time} presents a typical path with corresponding time allocations, as used in the performance study described in the Simulations Section. 

\begin{table}[H]  %
\centering        
\caption{\textbf{Cyclic Annealing Path with Time}}  %
\begin{tabular}{c|c}  
\hline
time ($\mu s$) & $(b_z, b_x)$ \\ \hline
$0$ & $(0,0)$ \\ \hline
$0.1$ & $(0.03,0)$ \\ \hline
$0.6$ & $(0.03,8.04)$ \\ \hline
$300$ & $(0,0)$ \\ \hline
\end{tabular}
\label{table:path-time}
\end{table}

During the first step of the algorithm, all eigenstates are bitstring states, as the non-commuting driving Hamiltonian (Eq.~\ref{eq:driving}) is not yet turned on. The evolution in this part can be fast. The second step, characterized by a finite energy gap above the ground state, can also be executed rapidly without consuming substantial annealing time. However, the third step, which goes into the spin-glass phase and encounters numerous Landau-Zener transitions, necessitates a slower pace. It is recommended to set the annealing time within the range of 100 to 1000 $\mu$s. In our experiments, we used 300 $\mu$s in the Simulations section and 500 $\mu$s in the subsequent sections.

It is noteworthy that the cycle's shape is not confined to a triangular or trapezoidal shape. An intriguing and open question remains on how different cycle shapes influence the annealing performance and the quest for an optimal one.

\section{acknowledgments}
We are grateful to Mohammad Amin, Kevin Chern, Pau Farré, Fiona Hanington, Emile Hoskinson, Andrew King, Jack Raymond, Hanteng Wang, and Hsiu-Chung Yeh for useful discussions. This work was supported  by the NSF grants DMR-2037654 and DMR-2338819.

\section{Data availability}
All data analyzed in this study are available from the corresponding author upon reasonable request.


\bibliography{sn-bibliography}

\begin{thebibliography}{54}%
\makeatletter
\providecommand \@ifxundefined [1]{%
 \@ifx{#1\undefined}
}%
\providecommand \@ifnum [1]{%
 \ifnum #1\expandafter \@firstoftwo
 \else \expandafter \@secondoftwo
 \fi
}%
\providecommand \@ifx [1]{%
 \ifx #1\expandafter \@firstoftwo
 \else \expandafter \@secondoftwo
 \fi
}%
\providecommand \natexlab [1]{#1}%
\providecommand \enquote  [1]{``#1''}%
\providecommand \bibnamefont  [1]{#1}%
\providecommand \bibfnamefont [1]{#1}%
\providecommand \citenamefont [1]{#1}%
\providecommand \href@noop [0]{\@secondoftwo}%
\providecommand \href [0]{\begingroup \@sanitize@url \@href}%
\providecommand \@href[1]{\@@startlink{#1}\@@href}%
\providecommand \@@href[1]{\endgroup#1\@@endlink}%
\providecommand \@sanitize@url [0]{\catcode `\\12\catcode `\$12\catcode `\&12\catcode `\#12\catcode `\^12\catcode `\_12\catcode `\%12\relax}%
\providecommand \@@startlink[1]{}%
\providecommand \@@endlink[0]{}%
\providecommand \url  [0]{\begingroup\@sanitize@url \@url }%
\providecommand \@url [1]{\endgroup\@href {#1}{\urlprefix }}%
\providecommand \urlprefix  [0]{URL }%
\providecommand \Eprint [0]{\href }%
\providecommand \doibase [0]{https://doi.org/}%
\providecommand \selectlanguage [0]{\@gobble}%
\providecommand \bibinfo  [0]{\@secondoftwo}%
\providecommand \bibfield  [0]{\@secondoftwo}%
\providecommand \translation [1]{[#1]}%
\providecommand \BibitemOpen [0]{}%
\providecommand \bibitemStop [0]{}%
\providecommand \bibitemNoStop [0]{.\EOS\space}%
\providecommand \EOS [0]{\spacefactor3000\relax}%
\providecommand \BibitemShut  [1]{\csname bibitem#1\endcsname}%
\let\auto@bib@innerbib\@empty
\bibitem [{\citenamefont {Feynman}(1982)}]{feynmanSimulatingPhysicsComputers1982a}%
  \BibitemOpen
  \bibfield  {author} {\bibinfo {author} {\bibfnamefont {R.~P.}\ \bibnamefont {Feynman}},\ }\bibfield  {title} {\bibinfo {title} {Simulating physics with computers},\ }\href {https://doi.org/10.1007/BF02650179} {\bibfield  {journal} {\bibinfo  {journal} {International Journal of Theoretical Physics}\ }\textbf {\bibinfo {volume} {21}},\ \bibinfo {pages} {467} (\bibinfo {year} {1982})}\BibitemShut {NoStop}%
\bibitem [{\citenamefont {Brooke}\ \emph {et~al.}(1999)\citenamefont {Brooke}, \citenamefont {Bitko}, \citenamefont {Rosenbaum},\ and\ \citenamefont {Aeppli}}]{brookeQuantumAnnealingDisordered1999b}%
  \BibitemOpen
  \bibfield  {author} {\bibinfo {author} {\bibfnamefont {J.}~\bibnamefont {Brooke}}, \bibinfo {author} {\bibfnamefont {D.}~\bibnamefont {Bitko}}, \bibinfo {author} {\bibfnamefont {T.~F.}\ \bibnamefont {Rosenbaum}},\ and\ \bibinfo {author} {\bibfnamefont {G.}~\bibnamefont {Aeppli}},\ }\bibfield  {title} {\bibinfo {title} {Quantum {{Annealing}} of a {{Disordered Magnet}}},\ }\href {https://doi.org/10.1126/science.284.5415.779} {\bibfield  {journal} {\bibinfo  {journal} {Science}\ }\textbf {\bibinfo {volume} {284}},\ \bibinfo {pages} {779} (\bibinfo {year} {1999})}\BibitemShut {NoStop}%
\bibitem [{\citenamefont {Johnson}\ \emph {et~al.}(2011)\citenamefont {Johnson}, \citenamefont {Amin}, \citenamefont {Gildert}, \citenamefont {Lanting}, \citenamefont {Hamze}, \citenamefont {Dickson}, \citenamefont {Harris}, \citenamefont {Berkley}, \citenamefont {Johansson}, \citenamefont {Bunyk}, \citenamefont {Chapple}, \citenamefont {Enderud}, \citenamefont {Hilton}, \citenamefont {Karimi}, \citenamefont {Ladizinsky}, \citenamefont {Ladizinsky}, \citenamefont {Oh}, \citenamefont {Perminov}, \citenamefont {Rich}, \citenamefont {Thom}, \citenamefont {Tolkacheva}, \citenamefont {Truncik}, \citenamefont {Uchaikin}, \citenamefont {Wang}, \citenamefont {Wilson},\ and\ \citenamefont {Rose}}]{johnsonQuantumAnnealingManufactured2011}%
  \BibitemOpen
  \bibfield  {author} {\bibinfo {author} {\bibfnamefont {M.~W.}\ \bibnamefont {Johnson}}, \bibinfo {author} {\bibfnamefont {M.~H.~S.}\ \bibnamefont {Amin}}, \bibinfo {author} {\bibfnamefont {S.}~\bibnamefont {Gildert}}, \bibinfo {author} {\bibfnamefont {T.}~\bibnamefont {Lanting}}, \bibinfo {author} {\bibfnamefont {F.}~\bibnamefont {Hamze}}, \bibinfo {author} {\bibfnamefont {N.}~\bibnamefont {Dickson}}, \bibinfo {author} {\bibfnamefont {R.}~\bibnamefont {Harris}}, \bibinfo {author} {\bibfnamefont {A.~J.}\ \bibnamefont {Berkley}}, \bibinfo {author} {\bibfnamefont {J.}~\bibnamefont {Johansson}}, \bibinfo {author} {\bibfnamefont {P.}~\bibnamefont {Bunyk}}, \bibinfo {author} {\bibfnamefont {E.~M.}\ \bibnamefont {Chapple}}, \bibinfo {author} {\bibfnamefont {C.}~\bibnamefont {Enderud}}, \bibinfo {author} {\bibfnamefont {J.~P.}\ \bibnamefont {Hilton}}, \bibinfo {author} {\bibfnamefont {K.}~\bibnamefont {Karimi}}, \bibinfo {author} {\bibfnamefont {E.}~\bibnamefont {Ladizinsky}}, \bibinfo {author} {\bibfnamefont
  {N.}~\bibnamefont {Ladizinsky}}, \bibinfo {author} {\bibfnamefont {T.}~\bibnamefont {Oh}}, \bibinfo {author} {\bibfnamefont {I.}~\bibnamefont {Perminov}}, \bibinfo {author} {\bibfnamefont {C.}~\bibnamefont {Rich}}, \bibinfo {author} {\bibfnamefont {M.~C.}\ \bibnamefont {Thom}}, \bibinfo {author} {\bibfnamefont {E.}~\bibnamefont {Tolkacheva}}, \bibinfo {author} {\bibfnamefont {C.~J.~S.}\ \bibnamefont {Truncik}}, \bibinfo {author} {\bibfnamefont {S.}~\bibnamefont {Uchaikin}}, \bibinfo {author} {\bibfnamefont {J.}~\bibnamefont {Wang}}, \bibinfo {author} {\bibfnamefont {B.}~\bibnamefont {Wilson}},\ and\ \bibinfo {author} {\bibfnamefont {G.}~\bibnamefont {Rose}},\ }\bibfield  {title} {\bibinfo {title} {Quantum annealing with manufactured spins},\ }\href {https://doi.org/10.1038/nature10012} {\bibfield  {journal} {\bibinfo  {journal} {Nature}\ }\textbf {\bibinfo {volume} {473}},\ \bibinfo {pages} {194} (\bibinfo {year} {2011})}\BibitemShut {NoStop}%
\bibitem [{\citenamefont {Dickson}\ \emph {et~al.}(2013)\citenamefont {Dickson}, \citenamefont {Johnson}, \citenamefont {Amin}, \citenamefont {Harris}, \citenamefont {Altomare}, \citenamefont {Berkley}, \citenamefont {Bunyk}, \citenamefont {Cai}, \citenamefont {Chapple}, \citenamefont {Chavez}, \citenamefont {Cioata}, \citenamefont {Cirip}, \citenamefont {Debuen}, \citenamefont {{Drew-Brook}}, \citenamefont {Enderud}, \citenamefont {Gildert}, \citenamefont {Hamze}, \citenamefont {Hilton}, \citenamefont {Hoskinson}, \citenamefont {Karimi}, \citenamefont {Ladizinsky}, \citenamefont {Ladizinsky}, \citenamefont {Lanting}, \citenamefont {Mahon}, \citenamefont {Neufeld}, \citenamefont {Oh}, \citenamefont {Perminov}, \citenamefont {Petroff}, \citenamefont {Przybysz}, \citenamefont {Rich}, \citenamefont {Spear}, \citenamefont {Tcaciuc}, \citenamefont {Thom}, \citenamefont {Tolkacheva}, \citenamefont {Uchaikin}, \citenamefont {Wang}, \citenamefont {Wilson}, \citenamefont {Merali},\ and\ \citenamefont
  {Rose}}]{dicksonThermallyAssistedQuantum2013a}%
  \BibitemOpen
  \bibfield  {author} {\bibinfo {author} {\bibfnamefont {N.~G.}\ \bibnamefont {Dickson}}, \bibinfo {author} {\bibfnamefont {M.~W.}\ \bibnamefont {Johnson}}, \bibinfo {author} {\bibfnamefont {M.~H.}\ \bibnamefont {Amin}}, \bibinfo {author} {\bibfnamefont {R.}~\bibnamefont {Harris}}, \bibinfo {author} {\bibfnamefont {F.}~\bibnamefont {Altomare}}, \bibinfo {author} {\bibfnamefont {A.~J.}\ \bibnamefont {Berkley}}, \bibinfo {author} {\bibfnamefont {P.}~\bibnamefont {Bunyk}}, \bibinfo {author} {\bibfnamefont {J.}~\bibnamefont {Cai}}, \bibinfo {author} {\bibfnamefont {E.~M.}\ \bibnamefont {Chapple}}, \bibinfo {author} {\bibfnamefont {P.}~\bibnamefont {Chavez}}, \bibinfo {author} {\bibfnamefont {F.}~\bibnamefont {Cioata}}, \bibinfo {author} {\bibfnamefont {T.}~\bibnamefont {Cirip}}, \bibinfo {author} {\bibfnamefont {P.}~\bibnamefont {Debuen}}, \bibinfo {author} {\bibfnamefont {M.}~\bibnamefont {{Drew-Brook}}}, \bibinfo {author} {\bibfnamefont {C.}~\bibnamefont {Enderud}}, \bibinfo {author} {\bibfnamefont
  {S.}~\bibnamefont {Gildert}}, \bibinfo {author} {\bibfnamefont {F.}~\bibnamefont {Hamze}}, \bibinfo {author} {\bibfnamefont {J.~P.}\ \bibnamefont {Hilton}}, \bibinfo {author} {\bibfnamefont {E.}~\bibnamefont {Hoskinson}}, \bibinfo {author} {\bibfnamefont {K.}~\bibnamefont {Karimi}}, \bibinfo {author} {\bibfnamefont {E.}~\bibnamefont {Ladizinsky}}, \bibinfo {author} {\bibfnamefont {N.}~\bibnamefont {Ladizinsky}}, \bibinfo {author} {\bibfnamefont {T.}~\bibnamefont {Lanting}}, \bibinfo {author} {\bibfnamefont {T.}~\bibnamefont {Mahon}}, \bibinfo {author} {\bibfnamefont {R.}~\bibnamefont {Neufeld}}, \bibinfo {author} {\bibfnamefont {T.}~\bibnamefont {Oh}}, \bibinfo {author} {\bibfnamefont {I.}~\bibnamefont {Perminov}}, \bibinfo {author} {\bibfnamefont {C.}~\bibnamefont {Petroff}}, \bibinfo {author} {\bibfnamefont {A.}~\bibnamefont {Przybysz}}, \bibinfo {author} {\bibfnamefont {C.}~\bibnamefont {Rich}}, \bibinfo {author} {\bibfnamefont {P.}~\bibnamefont {Spear}}, \bibinfo {author} {\bibfnamefont
  {A.}~\bibnamefont {Tcaciuc}}, \bibinfo {author} {\bibfnamefont {M.~C.}\ \bibnamefont {Thom}}, \bibinfo {author} {\bibfnamefont {E.}~\bibnamefont {Tolkacheva}}, \bibinfo {author} {\bibfnamefont {S.}~\bibnamefont {Uchaikin}}, \bibinfo {author} {\bibfnamefont {J.}~\bibnamefont {Wang}}, \bibinfo {author} {\bibfnamefont {A.~B.}\ \bibnamefont {Wilson}}, \bibinfo {author} {\bibfnamefont {Z.}~\bibnamefont {Merali}},\ and\ \bibinfo {author} {\bibfnamefont {G.}~\bibnamefont {Rose}},\ }\bibfield  {title} {\bibinfo {title} {Thermally assisted quantum annealing of a 16-qubit problem},\ }\href {https://doi.org/10.1038/ncomms2920} {\bibfield  {journal} {\bibinfo  {journal} {Nature Communications}\ }\textbf {\bibinfo {volume} {4}},\ \bibinfo {pages} {1903} (\bibinfo {year} {2013})}\BibitemShut {NoStop}%
\bibitem [{\citenamefont {King}\ \emph {et~al.}(2018)\citenamefont {King}, \citenamefont {Carrasquilla}, \citenamefont {Raymond}, \citenamefont {Ozfidan}, \citenamefont {Andriyash}, \citenamefont {Berkley}, \citenamefont {Reis}, \citenamefont {Lanting}, \citenamefont {Harris}, \citenamefont {Altomare}, \citenamefont {Boothby}, \citenamefont {Bunyk}, \citenamefont {Enderud}, \citenamefont {Fr{\'e}chette}, \citenamefont {Hoskinson}, \citenamefont {Ladizinsky}, \citenamefont {Oh}, \citenamefont {{Poulin-Lamarre}}, \citenamefont {Rich}, \citenamefont {Sato}, \citenamefont {Smirnov}, \citenamefont {Swenson}, \citenamefont {Volkmann}, \citenamefont {Whittaker}, \citenamefont {Yao}, \citenamefont {Ladizinsky}, \citenamefont {Johnson}, \citenamefont {Hilton},\ and\ \citenamefont {Amin}}]{kingObservationTopologicalPhenomena2018}%
  \BibitemOpen
  \bibfield  {author} {\bibinfo {author} {\bibfnamefont {A.~D.}\ \bibnamefont {King}}, \bibinfo {author} {\bibfnamefont {J.}~\bibnamefont {Carrasquilla}}, \bibinfo {author} {\bibfnamefont {J.}~\bibnamefont {Raymond}}, \bibinfo {author} {\bibfnamefont {I.}~\bibnamefont {Ozfidan}}, \bibinfo {author} {\bibfnamefont {E.}~\bibnamefont {Andriyash}}, \bibinfo {author} {\bibfnamefont {A.}~\bibnamefont {Berkley}}, \bibinfo {author} {\bibfnamefont {M.}~\bibnamefont {Reis}}, \bibinfo {author} {\bibfnamefont {T.}~\bibnamefont {Lanting}}, \bibinfo {author} {\bibfnamefont {R.}~\bibnamefont {Harris}}, \bibinfo {author} {\bibfnamefont {F.}~\bibnamefont {Altomare}}, \bibinfo {author} {\bibfnamefont {K.}~\bibnamefont {Boothby}}, \bibinfo {author} {\bibfnamefont {P.~I.}\ \bibnamefont {Bunyk}}, \bibinfo {author} {\bibfnamefont {C.}~\bibnamefont {Enderud}}, \bibinfo {author} {\bibfnamefont {A.}~\bibnamefont {Fr{\'e}chette}}, \bibinfo {author} {\bibfnamefont {E.}~\bibnamefont {Hoskinson}}, \bibinfo {author} {\bibfnamefont
  {N.}~\bibnamefont {Ladizinsky}}, \bibinfo {author} {\bibfnamefont {T.}~\bibnamefont {Oh}}, \bibinfo {author} {\bibfnamefont {G.}~\bibnamefont {{Poulin-Lamarre}}}, \bibinfo {author} {\bibfnamefont {C.}~\bibnamefont {Rich}}, \bibinfo {author} {\bibfnamefont {Y.}~\bibnamefont {Sato}}, \bibinfo {author} {\bibfnamefont {A.~Y.}\ \bibnamefont {Smirnov}}, \bibinfo {author} {\bibfnamefont {L.~J.}\ \bibnamefont {Swenson}}, \bibinfo {author} {\bibfnamefont {M.~H.}\ \bibnamefont {Volkmann}}, \bibinfo {author} {\bibfnamefont {J.}~\bibnamefont {Whittaker}}, \bibinfo {author} {\bibfnamefont {J.}~\bibnamefont {Yao}}, \bibinfo {author} {\bibfnamefont {E.}~\bibnamefont {Ladizinsky}}, \bibinfo {author} {\bibfnamefont {M.~W.}\ \bibnamefont {Johnson}}, \bibinfo {author} {\bibfnamefont {J.}~\bibnamefont {Hilton}},\ and\ \bibinfo {author} {\bibfnamefont {M.~H.}\ \bibnamefont {Amin}},\ }\bibfield  {title} {\bibinfo {title} {Observation of topological phenomena in a programmable lattice of 1,800 qubits},\ }\href
  {https://doi.org/10.1038/s41586-018-0410-x} {\bibfield  {journal} {\bibinfo  {journal} {Nature}\ }\textbf {\bibinfo {volume} {560}},\ \bibinfo {pages} {456} (\bibinfo {year} {2018})}\BibitemShut {NoStop}%
\bibitem [{\citenamefont {King}\ \emph {et~al.}(2022)\citenamefont {King}, \citenamefont {Suzuki}, \citenamefont {Raymond}, \citenamefont {Zucca}, \citenamefont {Lanting}, \citenamefont {Altomare}, \citenamefont {Berkley}, \citenamefont {Ejtemaee}, \citenamefont {Hoskinson}, \citenamefont {Huang}, \citenamefont {Ladizinsky}, \citenamefont {MacDonald}, \citenamefont {Marsden}, \citenamefont {Oh}, \citenamefont {{Poulin-Lamarre}}, \citenamefont {Reis}, \citenamefont {Rich}, \citenamefont {Sato}, \citenamefont {Whittaker}, \citenamefont {Yao}, \citenamefont {Harris}, \citenamefont {Lidar}, \citenamefont {Nishimori},\ and\ \citenamefont {Amin}}]{kingCoherentQuantumAnnealing2022}%
  \BibitemOpen
  \bibfield  {author} {\bibinfo {author} {\bibfnamefont {A.~D.}\ \bibnamefont {King}}, \bibinfo {author} {\bibfnamefont {S.}~\bibnamefont {Suzuki}}, \bibinfo {author} {\bibfnamefont {J.}~\bibnamefont {Raymond}}, \bibinfo {author} {\bibfnamefont {A.}~\bibnamefont {Zucca}}, \bibinfo {author} {\bibfnamefont {T.}~\bibnamefont {Lanting}}, \bibinfo {author} {\bibfnamefont {F.}~\bibnamefont {Altomare}}, \bibinfo {author} {\bibfnamefont {A.~J.}\ \bibnamefont {Berkley}}, \bibinfo {author} {\bibfnamefont {S.}~\bibnamefont {Ejtemaee}}, \bibinfo {author} {\bibfnamefont {E.}~\bibnamefont {Hoskinson}}, \bibinfo {author} {\bibfnamefont {S.}~\bibnamefont {Huang}}, \bibinfo {author} {\bibfnamefont {E.}~\bibnamefont {Ladizinsky}}, \bibinfo {author} {\bibfnamefont {A.~J.~R.}\ \bibnamefont {MacDonald}}, \bibinfo {author} {\bibfnamefont {G.}~\bibnamefont {Marsden}}, \bibinfo {author} {\bibfnamefont {T.}~\bibnamefont {Oh}}, \bibinfo {author} {\bibfnamefont {G.}~\bibnamefont {{Poulin-Lamarre}}}, \bibinfo {author} {\bibfnamefont
  {M.}~\bibnamefont {Reis}}, \bibinfo {author} {\bibfnamefont {C.}~\bibnamefont {Rich}}, \bibinfo {author} {\bibfnamefont {Y.}~\bibnamefont {Sato}}, \bibinfo {author} {\bibfnamefont {J.~D.}\ \bibnamefont {Whittaker}}, \bibinfo {author} {\bibfnamefont {J.}~\bibnamefont {Yao}}, \bibinfo {author} {\bibfnamefont {R.}~\bibnamefont {Harris}}, \bibinfo {author} {\bibfnamefont {D.~A.}\ \bibnamefont {Lidar}}, \bibinfo {author} {\bibfnamefont {H.}~\bibnamefont {Nishimori}},\ and\ \bibinfo {author} {\bibfnamefont {M.~H.}\ \bibnamefont {Amin}},\ }\bibfield  {title} {\bibinfo {title} {Coherent quantum annealing in a programmable 2,000 qubit {{Ising}} chain},\ }\href {https://doi.org/10.1038/s41567-022-01741-6} {\bibfield  {journal} {\bibinfo  {journal} {Nature Physics}\ }\textbf {\bibinfo {volume} {18}},\ \bibinfo {pages} {1324} (\bibinfo {year} {2022})}\BibitemShut {NoStop}%
\bibitem [{\citenamefont {Mohseni}\ \emph {et~al.}(2022)\citenamefont {Mohseni}, \citenamefont {McMahon},\ and\ \citenamefont {Byrnes}}]{mohseniIsingMachinesHardware2022}%
  \BibitemOpen
  \bibfield  {author} {\bibinfo {author} {\bibfnamefont {N.}~\bibnamefont {Mohseni}}, \bibinfo {author} {\bibfnamefont {P.~L.}\ \bibnamefont {McMahon}},\ and\ \bibinfo {author} {\bibfnamefont {T.}~\bibnamefont {Byrnes}},\ }\bibfield  {title} {\bibinfo {title} {Ising machines as hardware solvers of combinatorial optimization problems},\ }\href {https://doi.org/10.1038/s42254-022-00440-8} {\bibfield  {journal} {\bibinfo  {journal} {Nature Reviews Physics}\ }\textbf {\bibinfo {volume} {4}},\ \bibinfo {pages} {363} (\bibinfo {year} {2022})}\BibitemShut {NoStop}%
\bibitem [{\citenamefont {King}\ \emph {et~al.}(2023)\citenamefont {King}, \citenamefont {Raymond}, \citenamefont {Lanting}, \citenamefont {Harris}, \citenamefont {Zucca}, \citenamefont {Altomare}, \citenamefont {Berkley}, \citenamefont {Boothby}, \citenamefont {Ejtemaee}, \citenamefont {Enderud}, \citenamefont {Hoskinson}, \citenamefont {Huang}, \citenamefont {Ladizinsky}, \citenamefont {MacDonald}, \citenamefont {Marsden}, \citenamefont {Molavi}, \citenamefont {Oh}, \citenamefont {{Poulin-Lamarre}}, \citenamefont {Reis}, \citenamefont {Rich}, \citenamefont {Sato}, \citenamefont {Tsai}, \citenamefont {Volkmann}, \citenamefont {Whittaker}, \citenamefont {Yao}, \citenamefont {Sandvik},\ and\ \citenamefont {Amin}}]{kingQuantumCriticalDynamics2023}%
  \BibitemOpen
  \bibfield  {author} {\bibinfo {author} {\bibfnamefont {A.~D.}\ \bibnamefont {King}}, \bibinfo {author} {\bibfnamefont {J.}~\bibnamefont {Raymond}}, \bibinfo {author} {\bibfnamefont {T.}~\bibnamefont {Lanting}}, \bibinfo {author} {\bibfnamefont {R.}~\bibnamefont {Harris}}, \bibinfo {author} {\bibfnamefont {A.}~\bibnamefont {Zucca}}, \bibinfo {author} {\bibfnamefont {F.}~\bibnamefont {Altomare}}, \bibinfo {author} {\bibfnamefont {A.~J.}\ \bibnamefont {Berkley}}, \bibinfo {author} {\bibfnamefont {K.}~\bibnamefont {Boothby}}, \bibinfo {author} {\bibfnamefont {S.}~\bibnamefont {Ejtemaee}}, \bibinfo {author} {\bibfnamefont {C.}~\bibnamefont {Enderud}}, \bibinfo {author} {\bibfnamefont {E.}~\bibnamefont {Hoskinson}}, \bibinfo {author} {\bibfnamefont {S.}~\bibnamefont {Huang}}, \bibinfo {author} {\bibfnamefont {E.}~\bibnamefont {Ladizinsky}}, \bibinfo {author} {\bibfnamefont {A.~J.~R.}\ \bibnamefont {MacDonald}}, \bibinfo {author} {\bibfnamefont {G.}~\bibnamefont {Marsden}}, \bibinfo {author} {\bibfnamefont
  {R.}~\bibnamefont {Molavi}}, \bibinfo {author} {\bibfnamefont {T.}~\bibnamefont {Oh}}, \bibinfo {author} {\bibfnamefont {G.}~\bibnamefont {{Poulin-Lamarre}}}, \bibinfo {author} {\bibfnamefont {M.}~\bibnamefont {Reis}}, \bibinfo {author} {\bibfnamefont {C.}~\bibnamefont {Rich}}, \bibinfo {author} {\bibfnamefont {Y.}~\bibnamefont {Sato}}, \bibinfo {author} {\bibfnamefont {N.}~\bibnamefont {Tsai}}, \bibinfo {author} {\bibfnamefont {M.}~\bibnamefont {Volkmann}}, \bibinfo {author} {\bibfnamefont {J.~D.}\ \bibnamefont {Whittaker}}, \bibinfo {author} {\bibfnamefont {J.}~\bibnamefont {Yao}}, \bibinfo {author} {\bibfnamefont {A.~W.}\ \bibnamefont {Sandvik}},\ and\ \bibinfo {author} {\bibfnamefont {M.~H.}\ \bibnamefont {Amin}},\ }\bibfield  {title} {\bibinfo {title} {Quantum critical dynamics in a 5000-qubit programmable spin glass},\ }\bibfield  {journal} {\bibinfo  {journal} {Nature}\ }\href {https://doi.org/10.1038/s41586-023-05867-2} {10.1038/s41586-023-05867-2} (\bibinfo {year} {2023}),\ \Eprint
  {https://arxiv.org/abs/2207.13800} {arxiv:2207.13800 [cond-mat, physics:quant-ph]} \BibitemShut {NoStop}%
\bibitem [{\citenamefont {{Perdomo-Ortiz}}\ \emph {et~al.}(2012)\citenamefont {{Perdomo-Ortiz}}, \citenamefont {Dickson}, \citenamefont {{Drew-Brook}}, \citenamefont {Rose},\ and\ \citenamefont {{Aspuru-Guzik}}}]{perdomo-ortizFindingLowenergyConformations2012}%
  \BibitemOpen
  \bibfield  {author} {\bibinfo {author} {\bibfnamefont {A.}~\bibnamefont {{Perdomo-Ortiz}}}, \bibinfo {author} {\bibfnamefont {N.}~\bibnamefont {Dickson}}, \bibinfo {author} {\bibfnamefont {M.}~\bibnamefont {{Drew-Brook}}}, \bibinfo {author} {\bibfnamefont {G.}~\bibnamefont {Rose}},\ and\ \bibinfo {author} {\bibfnamefont {A.}~\bibnamefont {{Aspuru-Guzik}}},\ }\bibfield  {title} {\bibinfo {title} {Finding low-energy conformations of lattice protein models by quantum annealing},\ }\href {https://doi.org/10.1038/srep00571} {\bibfield  {journal} {\bibinfo  {journal} {Scientific Reports}\ }\textbf {\bibinfo {volume} {2}},\ \bibinfo {pages} {571} (\bibinfo {year} {2012})}\BibitemShut {NoStop}%
\bibitem [{\citenamefont {Harris}\ \emph {et~al.}(2018)\citenamefont {Harris}, \citenamefont {Sato}, \citenamefont {Berkley}, \citenamefont {Reis}, \citenamefont {Altomare}, \citenamefont {Amin}, \citenamefont {Boothby}, \citenamefont {Bunyk}, \citenamefont {Deng}, \citenamefont {Enderud}, \citenamefont {Huang}, \citenamefont {Hoskinson}, \citenamefont {Johnson}, \citenamefont {Ladizinsky}, \citenamefont {Ladizinsky}, \citenamefont {Lanting}, \citenamefont {Li}, \citenamefont {Medina}, \citenamefont {Molavi}, \citenamefont {Neufeld}, \citenamefont {Oh}, \citenamefont {Pavlov}, \citenamefont {Perminov}, \citenamefont {{Poulin-Lamarre}}, \citenamefont {Rich}, \citenamefont {Smirnov}, \citenamefont {Swenson}, \citenamefont {Tsai}, \citenamefont {Volkmann}, \citenamefont {Whittaker},\ and\ \citenamefont {Yao}}]{harrisPhaseTransitionsProgrammable2018}%
  \BibitemOpen
  \bibfield  {author} {\bibinfo {author} {\bibfnamefont {R.}~\bibnamefont {Harris}}, \bibinfo {author} {\bibfnamefont {Y.}~\bibnamefont {Sato}}, \bibinfo {author} {\bibfnamefont {A.~J.}\ \bibnamefont {Berkley}}, \bibinfo {author} {\bibfnamefont {M.}~\bibnamefont {Reis}}, \bibinfo {author} {\bibfnamefont {F.}~\bibnamefont {Altomare}}, \bibinfo {author} {\bibfnamefont {M.~H.}\ \bibnamefont {Amin}}, \bibinfo {author} {\bibfnamefont {K.}~\bibnamefont {Boothby}}, \bibinfo {author} {\bibfnamefont {P.}~\bibnamefont {Bunyk}}, \bibinfo {author} {\bibfnamefont {C.}~\bibnamefont {Deng}}, \bibinfo {author} {\bibfnamefont {C.}~\bibnamefont {Enderud}}, \bibinfo {author} {\bibfnamefont {S.}~\bibnamefont {Huang}}, \bibinfo {author} {\bibfnamefont {E.}~\bibnamefont {Hoskinson}}, \bibinfo {author} {\bibfnamefont {M.~W.}\ \bibnamefont {Johnson}}, \bibinfo {author} {\bibfnamefont {E.}~\bibnamefont {Ladizinsky}}, \bibinfo {author} {\bibfnamefont {N.}~\bibnamefont {Ladizinsky}}, \bibinfo {author} {\bibfnamefont {T.}~\bibnamefont
  {Lanting}}, \bibinfo {author} {\bibfnamefont {R.}~\bibnamefont {Li}}, \bibinfo {author} {\bibfnamefont {T.}~\bibnamefont {Medina}}, \bibinfo {author} {\bibfnamefont {R.}~\bibnamefont {Molavi}}, \bibinfo {author} {\bibfnamefont {R.}~\bibnamefont {Neufeld}}, \bibinfo {author} {\bibfnamefont {T.}~\bibnamefont {Oh}}, \bibinfo {author} {\bibfnamefont {I.}~\bibnamefont {Pavlov}}, \bibinfo {author} {\bibfnamefont {I.}~\bibnamefont {Perminov}}, \bibinfo {author} {\bibfnamefont {G.}~\bibnamefont {{Poulin-Lamarre}}}, \bibinfo {author} {\bibfnamefont {C.}~\bibnamefont {Rich}}, \bibinfo {author} {\bibfnamefont {A.}~\bibnamefont {Smirnov}}, \bibinfo {author} {\bibfnamefont {L.}~\bibnamefont {Swenson}}, \bibinfo {author} {\bibfnamefont {N.}~\bibnamefont {Tsai}}, \bibinfo {author} {\bibfnamefont {M.}~\bibnamefont {Volkmann}}, \bibinfo {author} {\bibfnamefont {J.}~\bibnamefont {Whittaker}},\ and\ \bibinfo {author} {\bibfnamefont {J.}~\bibnamefont {Yao}},\ }\bibfield  {title} {\bibinfo {title} {Phase transitions in a
  programmable quantum spin glass simulator},\ }\href {https://doi.org/10.1126/science.aat2025} {\bibfield  {journal} {\bibinfo  {journal} {Science}\ }\textbf {\bibinfo {volume} {361}},\ \bibinfo {pages} {162} (\bibinfo {year} {2018})}\BibitemShut {NoStop}%
\bibitem [{\citenamefont {Mott}\ \emph {et~al.}(2017)\citenamefont {Mott}, \citenamefont {Job}, \citenamefont {Vlimant}, \citenamefont {Lidar},\ and\ \citenamefont {Spiropulu}}]{mottSolvingHiggsOptimization2017b}%
  \BibitemOpen
  \bibfield  {author} {\bibinfo {author} {\bibfnamefont {A.}~\bibnamefont {Mott}}, \bibinfo {author} {\bibfnamefont {J.}~\bibnamefont {Job}}, \bibinfo {author} {\bibfnamefont {J.-R.}\ \bibnamefont {Vlimant}}, \bibinfo {author} {\bibfnamefont {D.}~\bibnamefont {Lidar}},\ and\ \bibinfo {author} {\bibfnamefont {M.}~\bibnamefont {Spiropulu}},\ }\bibfield  {title} {\bibinfo {title} {Solving a {{Higgs}} optimization problem with quantum annealing for machine learning},\ }\href {https://doi.org/10.1038/nature24047} {\bibfield  {journal} {\bibinfo  {journal} {Nature}\ }\textbf {\bibinfo {volume} {550}},\ \bibinfo {pages} {375} (\bibinfo {year} {2017})}\BibitemShut {NoStop}%
\bibitem [{\citenamefont {King}\ \emph {et~al.}(2021)\citenamefont {King}, \citenamefont {Batista}, \citenamefont {Raymond}, \citenamefont {Lanting}, \citenamefont {Ozfidan}, \citenamefont {{Poulin-Lamarre}}, \citenamefont {Zhang},\ and\ \citenamefont {Amin}}]{kingQuantumAnnealingSimulation2021}%
  \BibitemOpen
  \bibfield  {author} {\bibinfo {author} {\bibfnamefont {A.~D.}\ \bibnamefont {King}}, \bibinfo {author} {\bibfnamefont {C.~D.}\ \bibnamefont {Batista}}, \bibinfo {author} {\bibfnamefont {J.}~\bibnamefont {Raymond}}, \bibinfo {author} {\bibfnamefont {T.}~\bibnamefont {Lanting}}, \bibinfo {author} {\bibfnamefont {I.}~\bibnamefont {Ozfidan}}, \bibinfo {author} {\bibfnamefont {G.}~\bibnamefont {{Poulin-Lamarre}}}, \bibinfo {author} {\bibfnamefont {H.}~\bibnamefont {Zhang}},\ and\ \bibinfo {author} {\bibfnamefont {M.~H.}\ \bibnamefont {Amin}},\ }\bibfield  {title} {\bibinfo {title} {Quantum {{Annealing Simulation}} of {{Out-of-Equilibrium Magnetization}} in a {{Spin-Chain Compound}}},\ }\href {https://doi.org/10.1103/PRXQuantum.2.030317} {\bibfield  {journal} {\bibinfo  {journal} {PRX Quantum}\ }\textbf {\bibinfo {volume} {2}},\ \bibinfo {pages} {030317} (\bibinfo {year} {2021})}\BibitemShut {NoStop}%
\bibitem [{\citenamefont {Abel}\ and\ \citenamefont {Spannowsky}(2021)}]{abelQuantumFieldTheoreticSimulationPlatform2021}%
  \BibitemOpen
  \bibfield  {author} {\bibinfo {author} {\bibfnamefont {S.}~\bibnamefont {Abel}}\ and\ \bibinfo {author} {\bibfnamefont {M.}~\bibnamefont {Spannowsky}},\ }\bibfield  {title} {\bibinfo {title} {Quantum-{{Field-Theoretic Simulation Platform}} for {{Observing}} the {{Fate}} of the {{False Vacuum}}},\ }\href {https://doi.org/10.1103/PRXQuantum.2.010349} {\bibfield  {journal} {\bibinfo  {journal} {PRX Quantum}\ }\textbf {\bibinfo {volume} {2}},\ \bibinfo {pages} {010349} (\bibinfo {year} {2021})}\BibitemShut {NoStop}%
\bibitem [{\citenamefont {Barahona}(1982)}]{barahonaComputationalComplexityIsing1982}%
  \BibitemOpen
  \bibfield  {author} {\bibinfo {author} {\bibfnamefont {F.}~\bibnamefont {Barahona}},\ }\bibfield  {title} {\bibinfo {title} {On the computational complexity of {{Ising}} spin glass models},\ }\href {https://doi.org/10.1088/0305-4470/15/10/028} {\bibfield  {journal} {\bibinfo  {journal} {Journal of Physics A: Mathematical and General}\ }\textbf {\bibinfo {volume} {15}},\ \bibinfo {pages} {3241} (\bibinfo {year} {1982})}\BibitemShut {NoStop}%
\bibitem [{\citenamefont {Farhi}\ \emph {et~al.}(2001)\citenamefont {Farhi}, \citenamefont {Goldstone}, \citenamefont {Gutmann}, \citenamefont {Lapan}, \citenamefont {Lundgren},\ and\ \citenamefont {Preda}}]{farhiQuantumAdiabaticEvolution2001}%
  \BibitemOpen
  \bibfield  {author} {\bibinfo {author} {\bibfnamefont {E.}~\bibnamefont {Farhi}}, \bibinfo {author} {\bibfnamefont {J.}~\bibnamefont {Goldstone}}, \bibinfo {author} {\bibfnamefont {S.}~\bibnamefont {Gutmann}}, \bibinfo {author} {\bibfnamefont {J.}~\bibnamefont {Lapan}}, \bibinfo {author} {\bibfnamefont {A.}~\bibnamefont {Lundgren}},\ and\ \bibinfo {author} {\bibfnamefont {D.}~\bibnamefont {Preda}},\ }\bibfield  {title} {\bibinfo {title} {A {{Quantum Adiabatic Evolution Algorithm Applied}} to {{Random Instances}} of an {{NP-Complete Problem}}},\ }\href {https://doi.org/10.1126/science.1057726} {\bibfield  {journal} {\bibinfo  {journal} {Science}\ }\textbf {\bibinfo {volume} {292}},\ \bibinfo {pages} {472} (\bibinfo {year} {2001})}\BibitemShut {NoStop}%
\bibitem [{\citenamefont {Battaglia}\ \emph {et~al.}(2005)\citenamefont {Battaglia}, \citenamefont {Santoro},\ and\ \citenamefont {Tosatti}}]{battagliaOptimizationQuantumAnnealing2005}%
  \BibitemOpen
  \bibfield  {author} {\bibinfo {author} {\bibfnamefont {D.~A.}\ \bibnamefont {Battaglia}}, \bibinfo {author} {\bibfnamefont {G.~E.}\ \bibnamefont {Santoro}},\ and\ \bibinfo {author} {\bibfnamefont {E.}~\bibnamefont {Tosatti}},\ }\bibfield  {title} {\bibinfo {title} {Optimization by quantum annealing: {{Lessons}} from hard satisfiability problems},\ }\href {https://doi.org/10.1103/PhysRevE.71.066707} {\bibfield  {journal} {\bibinfo  {journal} {Physical Review E}\ }\textbf {\bibinfo {volume} {71}},\ \bibinfo {pages} {066707} (\bibinfo {year} {2005})}\BibitemShut {NoStop}%
\bibitem [{\citenamefont {Lucas}(2014)}]{lucasIsingFormulationsMany2014}%
  \BibitemOpen
  \bibfield  {author} {\bibinfo {author} {\bibfnamefont {A.}~\bibnamefont {Lucas}},\ }\bibfield  {title} {\bibinfo {title} {Ising formulations of many {{NP}} problems},\ }\href@noop {} {\bibfield  {journal} {\bibinfo  {journal} {Frontiers in Physics}\ }\textbf {\bibinfo {volume} {2}} (\bibinfo {year} {2014})}\BibitemShut {NoStop}%
\bibitem [{\citenamefont {Kirkpatrick}\ \emph {et~al.}(1983)\citenamefont {Kirkpatrick}, \citenamefont {Gelatt},\ and\ \citenamefont {Vecchi}}]{kirkpatrickOptimizationSimulatedAnnealing1983}%
  \BibitemOpen
  \bibfield  {author} {\bibinfo {author} {\bibfnamefont {S.}~\bibnamefont {Kirkpatrick}}, \bibinfo {author} {\bibfnamefont {C.~D.}\ \bibnamefont {Gelatt}},\ and\ \bibinfo {author} {\bibfnamefont {M.~P.}\ \bibnamefont {Vecchi}},\ }\bibfield  {title} {\bibinfo {title} {Optimization by {{Simulated Annealing}}},\ }\href {https://doi.org/10.1126/science.220.4598.671} {\bibfield  {journal} {\bibinfo  {journal} {Science}\ }\textbf {\bibinfo {volume} {220}},\ \bibinfo {pages} {671} (\bibinfo {year} {1983})}\BibitemShut {NoStop}%
\bibitem [{\citenamefont {Finnila}\ \emph {et~al.}(1994)\citenamefont {Finnila}, \citenamefont {Gomez}, \citenamefont {Sebenik}, \citenamefont {Stenson},\ and\ \citenamefont {Doll}}]{finnilaQuantumAnnealingNew1994}%
  \BibitemOpen
  \bibfield  {author} {\bibinfo {author} {\bibfnamefont {A.~B.}\ \bibnamefont {Finnila}}, \bibinfo {author} {\bibfnamefont {M.~A.}\ \bibnamefont {Gomez}}, \bibinfo {author} {\bibfnamefont {C.}~\bibnamefont {Sebenik}}, \bibinfo {author} {\bibfnamefont {C.}~\bibnamefont {Stenson}},\ and\ \bibinfo {author} {\bibfnamefont {J.~D.}\ \bibnamefont {Doll}},\ }\bibfield  {title} {\bibinfo {title} {Quantum annealing: {{A}} new method for minimizing multidimensional functions},\ }\href {https://doi.org/10.1016/0009-2614(94)00117-0} {\bibfield  {journal} {\bibinfo  {journal} {Chemical Physics Letters}\ }\textbf {\bibinfo {volume} {219}},\ \bibinfo {pages} {343} (\bibinfo {year} {1994})}\BibitemShut {NoStop}%
\bibitem [{\citenamefont {Kadowaki}\ and\ \citenamefont {Nishimori}(1998)}]{kadowakiQuantumAnnealingTransverse1998a}%
  \BibitemOpen
  \bibfield  {author} {\bibinfo {author} {\bibfnamefont {T.}~\bibnamefont {Kadowaki}}\ and\ \bibinfo {author} {\bibfnamefont {H.}~\bibnamefont {Nishimori}},\ }\bibfield  {title} {\bibinfo {title} {Quantum annealing in the transverse {{Ising}} model},\ }\href {https://doi.org/10.1103/PhysRevE.58.5355} {\bibfield  {journal} {\bibinfo  {journal} {Physical Review E}\ }\textbf {\bibinfo {volume} {58}},\ \bibinfo {pages} {5355} (\bibinfo {year} {1998})}\BibitemShut {NoStop}%
\bibitem [{\citenamefont {Santoro}\ \emph {et~al.}(2002)\citenamefont {Santoro}, \citenamefont {Marto{\v n}{\'a}k}, \citenamefont {Tosatti},\ and\ \citenamefont {Car}}]{santoroTheoryQuantumAnnealing2002a}%
  \BibitemOpen
  \bibfield  {author} {\bibinfo {author} {\bibfnamefont {G.~E.}\ \bibnamefont {Santoro}}, \bibinfo {author} {\bibfnamefont {R.}~\bibnamefont {Marto{\v n}{\'a}k}}, \bibinfo {author} {\bibfnamefont {E.}~\bibnamefont {Tosatti}},\ and\ \bibinfo {author} {\bibfnamefont {R.}~\bibnamefont {Car}},\ }\bibfield  {title} {\bibinfo {title} {Theory of {{Quantum Annealing}} of an {{Ising Spin Glass}}},\ }\href {https://doi.org/10.1126/science.1068774} {\bibfield  {journal} {\bibinfo  {journal} {Science}\ }\textbf {\bibinfo {volume} {295}},\ \bibinfo {pages} {2427} (\bibinfo {year} {2002})}\BibitemShut {NoStop}%
\bibitem [{\citenamefont {Das}\ and\ \citenamefont {Chakrabarti}(2008)}]{dasColloquiumQuantumAnnealing2008a}%
  \BibitemOpen
  \bibfield  {author} {\bibinfo {author} {\bibfnamefont {A.}~\bibnamefont {Das}}\ and\ \bibinfo {author} {\bibfnamefont {B.~K.}\ \bibnamefont {Chakrabarti}},\ }\bibfield  {title} {\bibinfo {title} {Colloquium: {{Quantum}} annealing and analog quantum computation},\ }\href {https://doi.org/10.1103/RevModPhys.80.1061} {\bibfield  {journal} {\bibinfo  {journal} {Reviews of Modern Physics}\ }\textbf {\bibinfo {volume} {80}},\ \bibinfo {pages} {1061} (\bibinfo {year} {2008})}\BibitemShut {NoStop}%
\bibitem [{\citenamefont {Morita}\ and\ \citenamefont {Nishimori}(2008)}]{moritaMathematicalFoundationQuantum2008}%
  \BibitemOpen
  \bibfield  {author} {\bibinfo {author} {\bibfnamefont {S.}~\bibnamefont {Morita}}\ and\ \bibinfo {author} {\bibfnamefont {H.}~\bibnamefont {Nishimori}},\ }\bibfield  {title} {\bibinfo {title} {Mathematical foundation of quantum annealing},\ }\href {https://doi.org/10.1063/1.2995837} {\bibfield  {journal} {\bibinfo  {journal} {Journal of Mathematical Physics}\ }\textbf {\bibinfo {volume} {49}},\ \bibinfo {pages} {125210} (\bibinfo {year} {2008})}\BibitemShut {NoStop}%
\bibitem [{\citenamefont {Young}\ \emph {et~al.}(2010)\citenamefont {Young}, \citenamefont {Knysh},\ and\ \citenamefont {Smelyanskiy}}]{youngFirstOrderPhaseTransition2010}%
  \BibitemOpen
  \bibfield  {author} {\bibinfo {author} {\bibfnamefont {A.~P.}\ \bibnamefont {Young}}, \bibinfo {author} {\bibfnamefont {S.}~\bibnamefont {Knysh}},\ and\ \bibinfo {author} {\bibfnamefont {V.~N.}\ \bibnamefont {Smelyanskiy}},\ }\bibfield  {title} {\bibinfo {title} {First-{{Order Phase Transition}} in the {{Quantum Adiabatic Algorithm}}},\ }\href {https://doi.org/10.1103/PhysRevLett.104.020502} {\bibfield  {journal} {\bibinfo  {journal} {Physical Review Letters}\ }\textbf {\bibinfo {volume} {104}},\ \bibinfo {pages} {020502} (\bibinfo {year} {2010})}\BibitemShut {NoStop}%
\bibitem [{\citenamefont {Albash}\ and\ \citenamefont {Lidar}(2018)}]{albashAdiabaticQuantumComputation2018b}%
  \BibitemOpen
  \bibfield  {author} {\bibinfo {author} {\bibfnamefont {T.}~\bibnamefont {Albash}}\ and\ \bibinfo {author} {\bibfnamefont {D.~A.}\ \bibnamefont {Lidar}},\ }\bibfield  {title} {\bibinfo {title} {Adiabatic quantum computation},\ }\href {https://doi.org/10.1103/RevModPhys.90.015002} {\bibfield  {journal} {\bibinfo  {journal} {Reviews of Modern Physics}\ }\textbf {\bibinfo {volume} {90}},\ \bibinfo {pages} {015002} (\bibinfo {year} {2018})}\BibitemShut {NoStop}%
\bibitem [{\citenamefont {Ohkuwa}\ \emph {et~al.}(2018)\citenamefont {Ohkuwa}, \citenamefont {Nishimori},\ and\ \citenamefont {Lidar}}]{ohkuwaReverseAnnealingFully2018a}%
  \BibitemOpen
  \bibfield  {author} {\bibinfo {author} {\bibfnamefont {M.}~\bibnamefont {Ohkuwa}}, \bibinfo {author} {\bibfnamefont {H.}~\bibnamefont {Nishimori}},\ and\ \bibinfo {author} {\bibfnamefont {D.~A.}\ \bibnamefont {Lidar}},\ }\bibfield  {title} {\bibinfo {title} {Reverse annealing for the fully connected \$p\$-spin model},\ }\href {https://doi.org/10.1103/PhysRevA.98.022314} {\bibfield  {journal} {\bibinfo  {journal} {Physical Review A}\ }\textbf {\bibinfo {volume} {98}},\ \bibinfo {pages} {022314} (\bibinfo {year} {2018})}\BibitemShut {NoStop}%
\bibitem [{\citenamefont {Yamashiro}\ \emph {et~al.}(2019)\citenamefont {Yamashiro}, \citenamefont {Ohkuwa}, \citenamefont {Nishimori},\ and\ \citenamefont {Lidar}}]{yamashiroDynamicsReverseAnnealing2019}%
  \BibitemOpen
  \bibfield  {author} {\bibinfo {author} {\bibfnamefont {Y.}~\bibnamefont {Yamashiro}}, \bibinfo {author} {\bibfnamefont {M.}~\bibnamefont {Ohkuwa}}, \bibinfo {author} {\bibfnamefont {H.}~\bibnamefont {Nishimori}},\ and\ \bibinfo {author} {\bibfnamefont {D.~A.}\ \bibnamefont {Lidar}},\ }\bibfield  {title} {\bibinfo {title} {Dynamics of reverse annealing for the fully connected \$p\$-spin model},\ }\href {https://doi.org/10.1103/PhysRevA.100.052321} {\bibfield  {journal} {\bibinfo  {journal} {Physical Review A}\ }\textbf {\bibinfo {volume} {100}},\ \bibinfo {pages} {052321} (\bibinfo {year} {2019})}\BibitemShut {NoStop}%
\bibitem [{\citenamefont {Chancellor}\ and\ \citenamefont {Kendon}()}]{chancellor_experimental_2021}%
  \BibitemOpen
  \bibfield  {author} {\bibinfo {author} {\bibfnamefont {N.}~\bibnamefont {Chancellor}}\ and\ \bibinfo {author} {\bibfnamefont {V.}~\bibnamefont {Kendon}},\ }\bibfield  {title} {\bibinfo {title} {Experimental test of search range in quantum annealing},\ }\href {https://doi.org/10.1103/PhysRevA.104.012604} {\bibfield  {journal} {\bibinfo  {journal} {Physical Review A}\ }\textbf {\bibinfo {volume} {104}},\ \bibinfo {pages} {012604}},\ \bibinfo {note} {publisher: American Physical Society}\BibitemShut {NoStop}%
\bibitem [{\citenamefont {Hauke}\ \emph {et~al.}(2020)\citenamefont {Hauke}, \citenamefont {Katzgraber}, \citenamefont {Lechner}, \citenamefont {Nishimori},\ and\ \citenamefont {Oliver}}]{haukePerspectivesQuantumAnnealing2020a}%
  \BibitemOpen
  \bibfield  {author} {\bibinfo {author} {\bibfnamefont {P.}~\bibnamefont {Hauke}}, \bibinfo {author} {\bibfnamefont {H.~G.}\ \bibnamefont {Katzgraber}}, \bibinfo {author} {\bibfnamefont {W.}~\bibnamefont {Lechner}}, \bibinfo {author} {\bibfnamefont {H.}~\bibnamefont {Nishimori}},\ and\ \bibinfo {author} {\bibfnamefont {W.~D.}\ \bibnamefont {Oliver}},\ }\bibfield  {title} {\bibinfo {title} {Perspectives of quantum annealing: Methods and implementations},\ }\href {https://doi.org/10.1088/1361-6633/ab85b8} {\bibfield  {journal} {\bibinfo  {journal} {Reports on Progress in Physics}\ }\textbf {\bibinfo {volume} {83}},\ \bibinfo {pages} {054401} (\bibinfo {year} {2020})}\BibitemShut {NoStop}%
\bibitem [{\citenamefont {Passarelli}\ \emph {et~al.}(2020)\citenamefont {Passarelli}, \citenamefont {Yip}, \citenamefont {Lidar}, \citenamefont {Nishimori},\ and\ \citenamefont {Lucignano}}]{passarelliReverseQuantumAnnealing2020}%
  \BibitemOpen
  \bibfield  {author} {\bibinfo {author} {\bibfnamefont {G.}~\bibnamefont {Passarelli}}, \bibinfo {author} {\bibfnamefont {K.-W.}\ \bibnamefont {Yip}}, \bibinfo {author} {\bibfnamefont {D.~A.}\ \bibnamefont {Lidar}}, \bibinfo {author} {\bibfnamefont {H.}~\bibnamefont {Nishimori}},\ and\ \bibinfo {author} {\bibfnamefont {P.}~\bibnamefont {Lucignano}},\ }\bibfield  {title} {\bibinfo {title} {Reverse quantum annealing of the \$p\$-spin model with relaxation},\ }\href {https://doi.org/10.1103/PhysRevA.101.022331} {\bibfield  {journal} {\bibinfo  {journal} {Physical Review A}\ }\textbf {\bibinfo {volume} {101}},\ \bibinfo {pages} {022331} (\bibinfo {year} {2020})}\BibitemShut {NoStop}%
\bibitem [{\citenamefont {Rajak}\ \emph {et~al.}(2023)\citenamefont {Rajak}, \citenamefont {Suzuki}, \citenamefont {Dutta},\ and\ \citenamefont {Chakrabarti}}]{rajakQuantumAnnealingOverview2023}%
  \BibitemOpen
  \bibfield  {author} {\bibinfo {author} {\bibfnamefont {A.}~\bibnamefont {Rajak}}, \bibinfo {author} {\bibfnamefont {S.}~\bibnamefont {Suzuki}}, \bibinfo {author} {\bibfnamefont {A.}~\bibnamefont {Dutta}},\ and\ \bibinfo {author} {\bibfnamefont {B.~K.}\ \bibnamefont {Chakrabarti}},\ }\bibfield  {title} {\bibinfo {title} {Quantum {{Annealing}}: {{An Overview}}},\ }\href {https://doi.org/10.1098/rsta.2021.0417} {\bibfield  {journal} {\bibinfo  {journal} {Philosophical Transactions of the Royal Society A: Mathematical, Physical and Engineering Sciences}\ }\textbf {\bibinfo {volume} {381}},\ \bibinfo {pages} {20210417} (\bibinfo {year} {2023})},\ \Eprint {https://arxiv.org/abs/2207.01827} {arxiv:2207.01827 [cond-mat, physics:quant-ph]} \BibitemShut {NoStop}%
\bibitem [{\citenamefont {Santoro}\ and\ \citenamefont {Tosatti}(2006)}]{santoroTOPICALREVIEWOptimization2006}%
  \BibitemOpen
  \bibfield  {author} {\bibinfo {author} {\bibfnamefont {G.~E.}\ \bibnamefont {Santoro}}\ and\ \bibinfo {author} {\bibfnamefont {E.}~\bibnamefont {Tosatti}},\ }\bibfield  {title} {\bibinfo {title} {{{TOPICAL REVIEW}}: {{Optimization}} using quantum mechanics: Quantum annealing through adiabatic evolution},\ }\href {https://doi.org/10.1088/0305-4470/39/36/R01} {\bibfield  {journal} {\bibinfo  {journal} {Journal of Physics A Mathematical General}\ }\textbf {\bibinfo {volume} {39}},\ \bibinfo {pages} {R393} (\bibinfo {year} {2006})}\BibitemShut {NoStop}%
\bibitem [{\citenamefont {Edwards}\ and\ \citenamefont {Anderson}(1975)}]{edwardsTheorySpinGlasses1975}%
  \BibitemOpen
  \bibfield  {author} {\bibinfo {author} {\bibfnamefont {S.~F.}\ \bibnamefont {Edwards}}\ and\ \bibinfo {author} {\bibfnamefont {P.~W.}\ \bibnamefont {Anderson}},\ }\bibfield  {title} {\bibinfo {title} {Theory of spin glasses},\ }\href {https://doi.org/10.1088/0305-4608/5/5/017} {\bibfield  {journal} {\bibinfo  {journal} {Journal of Physics F Metal Physics}\ }\textbf {\bibinfo {volume} {5}},\ \bibinfo {pages} {965} (\bibinfo {year} {1975})}\BibitemShut {NoStop}%
\bibitem [{\citenamefont {Binder}\ and\ \citenamefont {Young}(1986)}]{binderSpinGlassesExperimental1986a}%
  \BibitemOpen
  \bibfield  {author} {\bibinfo {author} {\bibfnamefont {K.}~\bibnamefont {Binder}}\ and\ \bibinfo {author} {\bibfnamefont {A.~P.}\ \bibnamefont {Young}},\ }\bibfield  {title} {\bibinfo {title} {Spin glasses: {{Experimental}} facts, theoretical concepts, and open questions},\ }\href {https://doi.org/10.1103/RevModPhys.58.801} {\bibfield  {journal} {\bibinfo  {journal} {Reviews of Modern Physics}\ }\textbf {\bibinfo {volume} {58}},\ \bibinfo {pages} {801} (\bibinfo {year} {1986})}\BibitemShut {NoStop}%
\bibitem [{\citenamefont {Crisanti}\ \emph {et~al.}(2003)\citenamefont {Crisanti}, \citenamefont {Leuzzi}, \citenamefont {Parisi},\ and\ \citenamefont {Rizzo}}]{crisantiComplexitySherringtonKirkpatrickModel2003}%
  \BibitemOpen
  \bibfield  {author} {\bibinfo {author} {\bibfnamefont {A.}~\bibnamefont {Crisanti}}, \bibinfo {author} {\bibfnamefont {L.}~\bibnamefont {Leuzzi}}, \bibinfo {author} {\bibfnamefont {G.}~\bibnamefont {Parisi}},\ and\ \bibinfo {author} {\bibfnamefont {T.}~\bibnamefont {Rizzo}},\ }\bibfield  {title} {\bibinfo {title} {Complexity in the {{Sherrington-Kirkpatrick}} model in the annealed approximation},\ }\href {https://doi.org/10.1103/PhysRevB.68.174401} {\bibfield  {journal} {\bibinfo  {journal} {Physical Review B}\ }\textbf {\bibinfo {volume} {68}},\ \bibinfo {pages} {174401} (\bibinfo {year} {2003})}\BibitemShut {NoStop}%
\bibitem [{\citenamefont {Cavagna}\ \emph {et~al.}(2004)\citenamefont {Cavagna}, \citenamefont {Giardina},\ and\ \citenamefont {Parisi}}]{cavagnaNumericalStudyMetastable2004}%
  \BibitemOpen
  \bibfield  {author} {\bibinfo {author} {\bibfnamefont {A.}~\bibnamefont {Cavagna}}, \bibinfo {author} {\bibfnamefont {I.}~\bibnamefont {Giardina}},\ and\ \bibinfo {author} {\bibfnamefont {G.}~\bibnamefont {Parisi}},\ }\bibfield  {title} {\bibinfo {title} {Numerical study of metastable states in {{Ising}} spin glasses},\ }\href {https://doi.org/10.1103/PhysRevLett.92.120603} {\bibfield  {journal} {\bibinfo  {journal} {Physical Review Letters}\ }\textbf {\bibinfo {volume} {92}},\ \bibinfo {pages} {120603} (\bibinfo {year} {2004})},\ \Eprint {https://arxiv.org/abs/cond-mat/0312534} {arxiv:cond-mat/0312534} \BibitemShut {NoStop}%
\bibitem [{\citenamefont {Mukherjee}\ \emph {et~al.}(2018{\natexlab{a}})\citenamefont {Mukherjee}, \citenamefont {Rajak},\ and\ \citenamefont {Chakrabarti}}]{mukherjeePossibleErgodicnonergodicRegions2018}%
  \BibitemOpen
  \bibfield  {author} {\bibinfo {author} {\bibfnamefont {S.}~\bibnamefont {Mukherjee}}, \bibinfo {author} {\bibfnamefont {A.}~\bibnamefont {Rajak}},\ and\ \bibinfo {author} {\bibfnamefont {B.~K.}\ \bibnamefont {Chakrabarti}},\ }\bibfield  {title} {\bibinfo {title} {Possible ergodic-nonergodic regions in the quantum {{Sherrington-Kirkpatrick}} spin glass model and quantum annealing},\ }\href {https://doi.org/10.1103/PhysRevE.97.022146} {\bibfield  {journal} {\bibinfo  {journal} {Physical Review E}\ }\textbf {\bibinfo {volume} {97}},\ \bibinfo {pages} {022146} (\bibinfo {year} {2018}{\natexlab{a}})}\BibitemShut {NoStop}%
\bibitem [{\citenamefont {Zener}\ and\ \citenamefont {Fowler}(1932)}]{zenerNonadiabaticCrossingEnergy1932}%
  \BibitemOpen
  \bibfield  {author} {\bibinfo {author} {\bibfnamefont {C.}~\bibnamefont {Zener}}\ and\ \bibinfo {author} {\bibfnamefont {R.~H.}\ \bibnamefont {Fowler}},\ }\bibfield  {title} {\bibinfo {title} {Non-adiabatic crossing of energy levels},\ }\href {https://doi.org/10.1098/rspa.1932.0165} {\bibfield  {journal} {\bibinfo  {journal} {Proceedings of the Royal Society of London. Series A, Containing Papers of a Mathematical and Physical Character}\ }\textbf {\bibinfo {volume} {137}},\ \bibinfo {pages} {696} (\bibinfo {year} {1932})}\BibitemShut {NoStop}%
\bibitem [{\citenamefont {Sinitsyn}(2002)}]{sinitsynMultiparticleLandauZenerProblem2002}%
  \BibitemOpen
  \bibfield  {author} {\bibinfo {author} {\bibfnamefont {N.~A.}\ \bibnamefont {Sinitsyn}},\ }\bibfield  {title} {\bibinfo {title} {Multiparticle {{Landau-Zener}} problem: {{Application}} to quantum dots},\ }\href {https://doi.org/10.1103/PhysRevB.66.205303} {\bibfield  {journal} {\bibinfo  {journal} {Physical Review B}\ }\textbf {\bibinfo {volume} {66}},\ \bibinfo {pages} {205303} (\bibinfo {year} {2002})}\BibitemShut {NoStop}%
\bibitem [{\citenamefont {Volkov}\ and\ \citenamefont {Ostrovsky}(2004)}]{volkovExactResultsSurvival2004}%
  \BibitemOpen
  \bibfield  {author} {\bibinfo {author} {\bibfnamefont {M.~V.}\ \bibnamefont {Volkov}}\ and\ \bibinfo {author} {\bibfnamefont {V.~N.}\ \bibnamefont {Ostrovsky}},\ }\bibfield  {title} {\bibinfo {title} {Exact results for survival probability in the multistate {{Landau}}{\textendash}{{Zener}} model},\ }\href {https://doi.org/10.1088/0953-4075/37/20/003} {\bibfield  {journal} {\bibinfo  {journal} {Journal of Physics B: Atomic, Molecular and Optical Physics}\ }\textbf {\bibinfo {volume} {37}},\ \bibinfo {pages} {4069} (\bibinfo {year} {2004})}\BibitemShut {NoStop}%
\bibitem [{\citenamefont {Damski}(2005)}]{damskiSimplestQuantumModel2005}%
  \BibitemOpen
  \bibfield  {author} {\bibinfo {author} {\bibfnamefont {B.}~\bibnamefont {Damski}},\ }\bibfield  {title} {\bibinfo {title} {The {{Simplest Quantum Model Supporting}} the {{Kibble-Zurek Mechanism}} of {{Topological Defect Production}}: {{Landau-Zener Transitions}} from a {{New Perspective}}},\ }\href {https://doi.org/10.1103/PhysRevLett.95.035701} {\bibfield  {journal} {\bibinfo  {journal} {Physical Review Letters}\ }\textbf {\bibinfo {volume} {95}},\ \bibinfo {pages} {035701} (\bibinfo {year} {2005})}\BibitemShut {NoStop}%
\bibitem [{\citenamefont {Sinitsyn}\ and\ \citenamefont {Li}(2016)}]{sinitsynSolvableMultistateModel2016}%
  \BibitemOpen
  \bibfield  {author} {\bibinfo {author} {\bibfnamefont {N.~A.}\ \bibnamefont {Sinitsyn}}\ and\ \bibinfo {author} {\bibfnamefont {F.}~\bibnamefont {Li}},\ }\bibfield  {title} {\bibinfo {title} {Solvable multistate model of {{Landau-Zener}} transitions in cavity {{QED}}},\ }\href {https://doi.org/10.1103/PhysRevA.93.063859} {\bibfield  {journal} {\bibinfo  {journal} {Physical Review A}\ }\textbf {\bibinfo {volume} {93}},\ \bibinfo {pages} {063859} (\bibinfo {year} {2016})}\BibitemShut {NoStop}%
\bibitem [{\citenamefont {Wang}\ \emph {et~al.}(2022)\citenamefont {Wang}, \citenamefont {Yeh},\ and\ \citenamefont {Kamenev}}]{wangManybodyLocalizationEnables2022}%
  \BibitemOpen
  \bibfield  {author} {\bibinfo {author} {\bibfnamefont {H.}~\bibnamefont {Wang}}, \bibinfo {author} {\bibfnamefont {H.-C.}\ \bibnamefont {Yeh}},\ and\ \bibinfo {author} {\bibfnamefont {A.}~\bibnamefont {Kamenev}},\ }\bibfield  {title} {\bibinfo {title} {Many-body localization enables iterative quantum optimization},\ }\href {https://doi.org/10.1038/s41467-022-33179-y} {\bibfield  {journal} {\bibinfo  {journal} {Nature Communications}\ }\textbf {\bibinfo {volume} {13}},\ \bibinfo {pages} {5503} (\bibinfo {year} {2022})}\BibitemShut {NoStop}%
\bibitem [{\citenamefont {Perdomo}\ \emph {et~al.}(2010)\citenamefont {Perdomo}, \citenamefont {{Venegas-Andraca}},\ and\ \citenamefont {{Aspuru-Guzik}}}]{perdomoStudyHeuristicGuesses2010}%
  \BibitemOpen
  \bibfield  {author} {\bibinfo {author} {\bibfnamefont {A.}~\bibnamefont {Perdomo}}, \bibinfo {author} {\bibfnamefont {S.~E.}\ \bibnamefont {{Venegas-Andraca}}},\ and\ \bibinfo {author} {\bibfnamefont {A.}~\bibnamefont {{Aspuru-Guzik}}},\ }\href {https://doi.org/10.48550/arXiv.0807.0354} {\bibinfo {title} {A study of heuristic guesses for adiabatic quantum computation}} (\bibinfo {year} {2010}),\ \Eprint {https://arxiv.org/abs/0807.0354} {arxiv:0807.0354 [quant-ph]} \BibitemShut {NoStop}%
\bibitem [{\citenamefont {Chancellor}(2017)}]{chancellorModernizingQuantumAnnealing2017}%
  \BibitemOpen
  \bibfield  {author} {\bibinfo {author} {\bibfnamefont {N.}~\bibnamefont {Chancellor}},\ }\bibfield  {title} {\bibinfo {title} {Modernizing quantum annealing using local searches},\ }\href {https://doi.org/10.1088/1367-2630/aa59c4} {\bibfield  {journal} {\bibinfo  {journal} {New Journal of Physics}\ }\textbf {\bibinfo {volume} {19}},\ \bibinfo {pages} {023024} (\bibinfo {year} {2017})}\BibitemShut {NoStop}%
\bibitem [{\citenamefont {Cao}\ \emph {et~al.}(2021)\citenamefont {Cao}, \citenamefont {Xue}, \citenamefont {Shannon},\ and\ \citenamefont {Joynt}}]{caoSpeedupQuantumAdiabatic2021}%
  \BibitemOpen
  \bibfield  {author} {\bibinfo {author} {\bibfnamefont {C.}~\bibnamefont {Cao}}, \bibinfo {author} {\bibfnamefont {J.}~\bibnamefont {Xue}}, \bibinfo {author} {\bibfnamefont {N.}~\bibnamefont {Shannon}},\ and\ \bibinfo {author} {\bibfnamefont {R.}~\bibnamefont {Joynt}},\ }\bibfield  {title} {\bibinfo {title} {Speedup of the quantum adiabatic algorithm using delocalization catalysis},\ }\href {https://doi.org/10.1103/PhysRevResearch.3.013092} {\bibfield  {journal} {\bibinfo  {journal} {Physical Review Research}\ }\textbf {\bibinfo {volume} {3}},\ \bibinfo {pages} {013092} (\bibinfo {year} {2021})}\BibitemShut {NoStop}%
\bibitem [{\citenamefont {Pal}\ and\ \citenamefont {Huse}(2010)}]{palManybodyLocalizationPhase2010a}%
  \BibitemOpen
  \bibfield  {author} {\bibinfo {author} {\bibfnamefont {A.}~\bibnamefont {Pal}}\ and\ \bibinfo {author} {\bibfnamefont {D.~A.}\ \bibnamefont {Huse}},\ }\bibfield  {title} {\bibinfo {title} {Many-body localization phase transition},\ }\href {https://doi.org/10.1103/PhysRevB.82.174411} {\bibfield  {journal} {\bibinfo  {journal} {Physical Review B}\ }\textbf {\bibinfo {volume} {82}},\ \bibinfo {pages} {174411} (\bibinfo {year} {2010})}\BibitemShut {NoStop}%
\bibitem [{\citenamefont {Gornyi}\ \emph {et~al.}(2016)\citenamefont {Gornyi}, \citenamefont {Mirlin},\ and\ \citenamefont {Polyakov}}]{gornyiManybodyDelocalizationTransition2016}%
  \BibitemOpen
  \bibfield  {author} {\bibinfo {author} {\bibfnamefont {I.~V.}\ \bibnamefont {Gornyi}}, \bibinfo {author} {\bibfnamefont {A.~D.}\ \bibnamefont {Mirlin}},\ and\ \bibinfo {author} {\bibfnamefont {D.~G.}\ \bibnamefont {Polyakov}},\ }\bibfield  {title} {\bibinfo {title} {Many-body delocalization transition and relaxation in a quantum dot},\ }\href {https://doi.org/10.1103/PhysRevB.93.125419} {\bibfield  {journal} {\bibinfo  {journal} {Physical Review B}\ }\textbf {\bibinfo {volume} {93}},\ \bibinfo {pages} {125419} (\bibinfo {year} {2016})}\BibitemShut {NoStop}%
\bibitem [{\citenamefont {Mukherjee}\ \emph {et~al.}(2018{\natexlab{b}})\citenamefont {Mukherjee}, \citenamefont {Nag},\ and\ \citenamefont {Garg}}]{mukherjeeManybodyLocalizationdelocalizationTransition2018}%
  \BibitemOpen
  \bibfield  {author} {\bibinfo {author} {\bibfnamefont {S.}~\bibnamefont {Mukherjee}}, \bibinfo {author} {\bibfnamefont {S.}~\bibnamefont {Nag}},\ and\ \bibinfo {author} {\bibfnamefont {A.}~\bibnamefont {Garg}},\ }\bibfield  {title} {\bibinfo {title} {Many-body localization-delocalization transition in the quantum {{Sherrington-Kirkpatrick}} model},\ }\href {https://doi.org/10.1103/PhysRevB.97.144202} {\bibfield  {journal} {\bibinfo  {journal} {Physical Review B}\ }\textbf {\bibinfo {volume} {97}},\ \bibinfo {pages} {144202} (\bibinfo {year} {2018}{\natexlab{b}})}\BibitemShut {NoStop}%
\bibitem [{\citenamefont {Palassini}\ and\ \citenamefont {Young}(1999)}]{palassiniTrivialityGroundState1999}%
  \BibitemOpen
  \bibfield  {author} {\bibinfo {author} {\bibfnamefont {M.}~\bibnamefont {Palassini}}\ and\ \bibinfo {author} {\bibfnamefont {A.~P.}\ \bibnamefont {Young}},\ }\bibfield  {title} {\bibinfo {title} {Triviality of the {{Ground State Structure}} in {{Ising Spin Glasses}}},\ }\href {https://doi.org/10.1103/PhysRevLett.83.5126} {\bibfield  {journal} {\bibinfo  {journal} {Physical Review Letters}\ }\textbf {\bibinfo {volume} {83}},\ \bibinfo {pages} {5126} (\bibinfo {year} {1999})},\ \Eprint {https://arxiv.org/abs/cond-mat/9906323} {arxiv:cond-mat/9906323} \BibitemShut {NoStop}%
\bibitem [{\citenamefont {Marinari}\ and\ \citenamefont {Parisi}(2000)}]{marinariEffectsChangingBoundary2000}%
  \BibitemOpen
  \bibfield  {author} {\bibinfo {author} {\bibfnamefont {E.}~\bibnamefont {Marinari}}\ and\ \bibinfo {author} {\bibfnamefont {G.}~\bibnamefont {Parisi}},\ }\bibfield  {title} {\bibinfo {title} {Effects of changing the boundary conditions on the ground state of {{Ising}} spin glasses},\ }\href {https://doi.org/10.1103/PhysRevB.62.11677} {\bibfield  {journal} {\bibinfo  {journal} {Physical Review B}\ }\textbf {\bibinfo {volume} {62}},\ \bibinfo {pages} {11677} (\bibinfo {year} {2000})}\BibitemShut {NoStop}%
\bibitem [{\citenamefont {Marinari}\ and\ \citenamefont {Parisi}(2001)}]{marinariEffectsBulkPerturbation2001}%
  \BibitemOpen
  \bibfield  {author} {\bibinfo {author} {\bibfnamefont {E.}~\bibnamefont {Marinari}}\ and\ \bibinfo {author} {\bibfnamefont {G.}~\bibnamefont {Parisi}},\ }\bibfield  {title} {\bibinfo {title} {Effects of a {{Bulk Perturbation}} on the {{Ground State}} of {{3D Ising Spin Glasses}}},\ }\href {https://doi.org/10.1103/PhysRevLett.86.3887} {\bibfield  {journal} {\bibinfo  {journal} {Physical Review Letters}\ }\textbf {\bibinfo {volume} {86}},\ \bibinfo {pages} {3887} (\bibinfo {year} {2001})}\BibitemShut {NoStop}%
\bibitem [{\citenamefont {Boothby}\ \emph {et~al.}()\citenamefont {Boothby}, \citenamefont {Bunyk}, \citenamefont {Raymond},\ and\ \citenamefont {Roy}}]{boothby_next-generation_2020}%
  \BibitemOpen
  \bibfield  {author} {\bibinfo {author} {\bibfnamefont {K.}~\bibnamefont {Boothby}}, \bibinfo {author} {\bibfnamefont {P.}~\bibnamefont {Bunyk}}, \bibinfo {author} {\bibfnamefont {J.}~\bibnamefont {Raymond}},\ and\ \bibinfo {author} {\bibfnamefont {A.}~\bibnamefont {Roy}},\ }\href {https://doi.org/10.48550/arXiv.2003.00133} {\bibinfo {title} {Next-generation topology of d-wave quantum processors}}\BibitemShut {NoStop}%
\bibitem [{\citenamefont {Misra-Spieldenner}\ \emph {et~al.}(2023)\citenamefont {Misra-Spieldenner}, \citenamefont {Bode}, \citenamefont {Schuhmacher}, \citenamefont {Stollenwerk}, \citenamefont {Bagrets},\ and\ \citenamefont {Wilhelm}}]{misra-spieldenner_mean-field_2023}%
  \BibitemOpen
  \bibfield  {author} {\bibinfo {author} {\bibfnamefont {A.}~\bibnamefont {Misra-Spieldenner}}, \bibinfo {author} {\bibfnamefont {T.}~\bibnamefont {Bode}}, \bibinfo {author} {\bibfnamefont {P.~K.}\ \bibnamefont {Schuhmacher}}, \bibinfo {author} {\bibfnamefont {T.}~\bibnamefont {Stollenwerk}}, \bibinfo {author} {\bibfnamefont {D.}~\bibnamefont {Bagrets}},\ and\ \bibinfo {author} {\bibfnamefont {F.~K.}\ \bibnamefont {Wilhelm}},\ }\bibfield  {title} {\bibinfo {title} {Mean-field approximate optimization algorithm},\ }\bibfield  {journal} {\bibinfo  {journal} {{PRX} Quantum}\ }\href {https://doi.org/10.1103/PRXQuantum.4.030335} {10.1103/PRXQuantum.4.030335} (\bibinfo {year} {2023})\BibitemShut {NoStop}%
\end{thebibliography}%

\end{document}